\documentclass[aps,eqsecnum,preprint,floats,epsf,epsfig,nofootinbib]{revtex4}
\usepackage{epsfig}

\begin{document}
\def\bea{\begin{eqnarray}}
\def\eea{\end{eqnarray}}
\def\nc{N_c^{\rm eff}}
\def\vp{\varepsilon}
\def\drho{\bar\rho}
\def\deta{\bar\eta}
\def\a{{\cal A}}
\def\B{{\cal B}}
\def\c{{\cal C}}
\def\d{{\cal D}}
\def\e{{\cal E}}
\def\p{{\cal P}}
\def\t{{\cal T}}
\def\up{\uparrow}
\def\dw{\downarrow}
\def\vma{{_{V-A}}}
\def\vpa{{_{V+A}}}
\def\smp{{_{S-P}}}
\def\spp{{_{S+P}}}
\def\J{{J/\psi}}
\def\ov{\overline}
\def\Lqcd{{\Lambda_{\rm QCD}}}
\def\pr{{Phys. Rev.}~}
\def\prl{{Phys. Rev. Lett.}~}
\def\pl{{Phys. Lett.}~}
\def\np{{Nucl. Phys.}~}
\def\zp{{Z. Phys.}~}
\def\lsim{ {\ \lower-1.2pt\vbox{\hbox{\rlap{$<$}\lower5pt\vbox{\hbox{$\sim$}
}}}\ } }
\def\gsim{ {\ \lower-1.2pt\vbox{\hbox{\rlap{$>$}\lower5pt\vbox{\hbox{$\sim$}
}}}\ } }

\font\el=cmbx10 scaled \magstep2{\obeylines\hfill June, 2005}

\vskip 1.5 cm
\begin{center}{\large \bf Annihilation
in Factorization-suppressed $B$ Decays Involving a $1^1P_1$ Meson
and Search for New-Physics Signals}
\end{center}

\bigskip
\centerline{\bf Kwei-Chou Yang} \centerline{Department of Physics,
Chung Yuan Christian University} \centerline{Chung-Li, Taiwan 320,
Republic of China}
\bigskip
\bigskip
\bigskip
\bigskip
\centerline{\bf Abstract}
\bigskip
The recent measurements of large transverse fractions in $B \to
\phi K^*$ decays represent a challenge for theory. It was shown
that the QCD penguin annihilation could contribute to the
transverse amplitudes such that the observation could be accounted
for in the Standard Model. However, if one resorts to new physics
for resolutions, then the relevant 4-quark operators should be
tensor-like or scalar-like. We show that the same annihilation
effects could remarkably enhance the longitudinal rates of
factorization-suppressed $\overline B^0\to h_1(1380) \overline
K^{*0}, b_1(1235) \overline K^{*0}$ modes to be
$(12.0^{+4.1}_{-3.0})\times 10^{-6}$ and $(7.0\pm 3.5)\times
10^{-6}$, respectively, but attend to be cancelled in the
transverse components. Nevertheless, the transverse fractions of
$B\to h_1(1380) K^*$ can become sizable due to new-physics
contributions. The measurements of these decays can thus help us
to realize the role of annihilation topologies in charmless $B$
decays and offer a theoretically clean window to search for the
evidence of new physics.

\pagebreak

\section{Introduction}
The BaBar and Belle collaborations have recently measured $B\to
\phi K^*$ decays with large transverse
fractions~\cite{Aubert:2003mm,Aubert:2004xc,Chen:2003jf,Abe:2004ku,hfag}
which are suppressed by $(1/m_b)^2$ in the Standard Model (SM)
perturbative picture. Within the SM, it has been given in
Ref.~\cite{Kagan} that the annihilation graphs, which are formally
suppressed by $1/m_b^2$ but logarithmically enhanced, could
account for the observations with a moderate value of the BBNS
parameter $\rho_A$~\cite{BBNS1}. However, the perturbative QCD
(PQCD) analysis~\cite{Li:2004ti} yielded the longitudinal fraction
$f_L \gsim 0.75$ even with considering the annihilation, in
contrast with the observation of $f_L\sim 0.5$. In the large
energy limit~\cite{Charles:1998dr}, since the three SM helicity
amplitudes of the $\phi K^*$ modes in the transversity basis are
respectively proportional to:
 \bea \label{eq:smamp}
{\overline A}_0^{SM} &\propto&  f_\phi m_B^2 \zeta_\parallel,
\nonumber \\
{\overline A}_\|^{SM} &\propto&  -{\sqrt 2} f_\phi m_\phi m_B
\zeta_\perp,
\nonumber \\
{\overline A}_\perp^{SM} &\propto& -{ \sqrt 2} f_\phi m_\phi m_B
\zeta_\perp,
 \end{eqnarray}
where
 \begin{eqnarray}
&&\frac{m_B}{m_B+m_{K^*}}V=\frac{m_B + m_{K^*}}{2E}A_1 =
\zeta_\perp\,,
\\
&& A_0=\frac{E}{m_{K^*}}\left(\frac{m_B + m_{K^*}}{2E}A_1
-\frac{m_B -m_{K^*}}{m_B}A_2  \right) =\zeta_\parallel\,,
 \end{eqnarray}
with $A_{0,1,2}$ and $V$ being the axial-vector and vector current
form factors, respectively, and the PQCD results for the $B \to
\phi K^*$ branching ratios (BRs) are about 1.5 times larger than
the data, it was thus suggested in Ref.~\cite{Li:2004mp} that
choosing a smaller $A_0$ could resolve the anomaly for observing
the large transverse fractions in the $\phi K^*$ modes. On the
other hand, it was argued
that~\cite{Colangelo:2004rd,Cheng:2004ru} the anomaly may be
resolved if the long distance final state interactions via charmed
meson intermediate states exist in the $B \to \phi K^*$ decays.

Some possible new physics (NP) solutions have been proposed. If
the mechanism is due to the right-handed currents, which could
contribute constructively to $\overline A_\perp$ but destructively
to $\overline A_{0,\|}$, then one may have larger $|\overline
A_\perp / \overline A_0 |^2$ to account for the data $f_L\sim
0.5$. Nevertheless, the resulting $|\overline A_{\|}|^2 \ll
|\overline A_\perp|^2$ will be in contrast to the recent
observations with $f_\perp\, {\rm (perpendicular\ fraction)} \sim
f_\parallel\, {\rm (parallel\
fraction)}$~\cite{Aubert:2004xc,Abe:2004ku}. (See further
explanations in Refs.~\cite{Kagan,Das:2004hq}.)  NP in $bsg$
chromomagnetic dipole operator was used to explain the large
transverse fractions of $\phi K^*$ modes in
Ref.~\cite{Hou:2004vj}. However, since, in large $m_b$ limit, the
strong interaction conserves the helicity of a produced light
quark pair, helicity conservation requires that the outgoing $s$
and $\bar s$ arising from $s-\bar s-n\ gluons$ vertex have {\it
opposite helicities}. The contribution of the chromomagnetic
dipole operator to the transverse polarization amplitudes should
be suppressed as $\overline H_{00}:\overline H_{--}: \overline
H_{++} \sim {\cal O}(1):{\cal O}(1/m_b): {\cal O} (1/m_b^2)$
\cite{Das:2004hq} which has no help for understanding the
observation. Furthermore, it has been shown in
Refs.~\cite{Kagan,Das:2004hq} that if considering only the two
parton scenario for the final mesons, the contributions of the
chromomagnetic dipole operator to the transverse polarization
amplitudes are actually equal to zero. An additional longitudinal
gluon is necessary for having non-vanishing transverse amplitudes.

A general discussion for searching possible NP solutions has been
given in \cite{Das:2004hq} that only two classes of NP four-quark
operators are relevant in resolving the transverse anomaly in the
$\phi K^*$ modes. The first class of operators with structures
$\sigma(1-\gamma_5)\otimes \sigma(1-\gamma_5)$ and
$(1-\gamma_5)\otimes (1-\gamma_5)$ contributes to helicity
amplitudes, which refer to the NP scenario 1, as ${\overline
H_{00}}:{\overline H_{--}}:{\overline H_{++}} \sim  {\cal
O}(1/m_b):{\cal O}(1/m_b^2):{\cal O}(1)$. The second class of
operators with structures $\sigma(1+\gamma_5)\otimes
\sigma(1+\gamma_5)$ and $(1+\gamma_5)\otimes (1+\gamma_5)$, which
refer to the NP scenario 2, results in ${\overline
H_{00}}:{\overline H_{++}}:{\overline H_{--}} \sim {\cal
O}(1/m_b):{\cal O}(1/m_b^2):{\cal O}(1)$~\footnote{For the $b\to s
\bar s s$ processes, the tensor operators can be transformed as
the scalar operators by Fierz transformation and vice
versa~\cite{Das:2004hq}. However it is not true for $b\to s \bar u
u$ and $b\to s \bar d d$. The tensor operators can be induced by
box diagrams with the exchange of two
gluinos~\cite{Borzumati:1999qt}.}. It was found that these two
classes can separately satisfy the two possible phase solutions
for polarization data, owing to the phase ambiguity in the
measurement, and resolve the anomaly for large transverse
fractions in the $\phi K^*$ modes. (Some discussions due to the
tensor operator $\sigma(1+\gamma_5)\otimes \sigma(1+\gamma_5)$,
can be found in \cite{Kagan,Kim:2004wq}.) A model application can
be found in Ref.~\cite{Chen:2005mk}.

In this paper, we shall devote to the study for
factorization-suppressed ${\overline B}\to V A$ decays, where $A\
(V)$ is an axial vector (vector) meson with quantum number
$N^{2S+1}L_j=1^1P_1\ (1^3S_1)$ in the quark model scenario. Some
$B$ decays involving $^1P_1$ mesons were discussed in
Ref.~\cite{Suzuki:2001za}. In particular, we focus on $h_1(1380)
K^*$ modes, where $h_1(1380)$ is a $1^1P_1$
meson\footnote{$h_1(1380)$ with $I^G(J^{PC})=?^-(1^{+-})$ was
denoted as $H'$ in old classification. Its isospin may be $0$, but
not confirmed yet.} and its properties are not well-established
experimentally~\cite{PDG}. The quark content of $h_1(1380)$ was
suggested as $\bar s s$ in the QCD sum rule
calculation~\cite{GRVW}. Due to the G-parity, the distribution
amplitudes of a $1^1P_1$ meson defined by the nonlocal vector and
axial-vector current are antisymmetric under the exchange of
$quark$ and $anti$-$quark$ momentum fractions in the SU(3) limit.
We shall show that in the SM, while the transverse components of
$h_1(1380) K^*$ modes are negligible, the longitudinal fraction
receiving large QCD corrections is further enhanced by the
annihilation topologies although it vanishes in the factorization
limit. Interestingly, the local tensor operator can couple mainly
to transversely polarized $h_1(1380)$ meson. This means that if
the large transverse fractions of $B\to \phi K^*$ decays are owing
to the NP 4-quark tensor operators, which contribute to the $b \to
s \bar s s $ processes, then we expect transverse branching
ratios: $ {\rm BR_T}(h_1(1380) K^*) \simeq {\rm BR_T}(\phi K^*)$
which would be striking evidence for physics beyond SM. We will
also show that the remarkable enhancement of the longitudinal
polarization due to the annihilation topologies could be observed
in $b_1(1235) K^*$ modes~\footnote{$b_1(1235)$ was denoted as
$B(1235)$ in old classification.}.

This paper is organized as follows. In Sec.~\ref{sec:h1ksamp}, we
begin with the summary of light-cone distribution amplitudes
(LCDAs) and introduce light-cone projection operator in the
momentum space that our QCD factorization results rely on. We then
calculate the QCD factorization decay amplitudes and take
$\overline B\to h_1(1380) \overline K^*$ as an example. In
Sec.~\ref{sec:np}, we give a detailed NP calculation for
$\overline B\to h_1(1380) K^*$, compared with $\overline B\to \phi
K^*$ results. Sec.~\ref{sec:results} contains numerical results
for several decay modes, along with a detailed estimation of
theoretical uncertainties from various sources. Finally, we
conclude in Sec.~\ref{sec:conclusion}.

\vspace{0.5cm} \noindent{\bf Brief summary of results}

Since Sec.~\ref{sec:h1ksamp} contains the mathematical expressions
for LCDAs of the $1^1P_1$ mesons and QCD factorization decay
amplitudes, and Sec.~\ref{sec:np} for new physics amplitudes, they
can be read independently. The reader, who is not familiar with
the theoretical framework, may temporarily omit these two sections
but consults Sec.~\ref{sec:results} about the numerical results,
for which we summarize the main branching ratios as follows. If
large transverse fractions in $B\to \phi K^*$ decays are owing to
the annihilation topology, we predict
 \begin{eqnarray}
 {\rm BR_{tot}}(h_1(1380)  K^{*0})&=&(12.0^{+4.1}_{-3.0})\times 10^{-6},\nonumber\\
 {\rm BR_{tot}}(b_1(1235)  K^{*0})&=&(7.0\pm 3.5)\times 10^{-6},
 \end{eqnarray}
which are predominated by the longitudinal component. Accordingly,
large ${\rm BR}(h_1(1380) K)$ and ${\rm BR} (b_1^+(1235) K^{-})$
could be observed. See also Sec.~\ref{sec:conclusion} for
discussions. On the other hand, if large transverse components of
the $\phi K^*$ modes originate from the NP, we can also observe
that large transverse fractions in $B\to h_1(1380) K^*$, such that
 \begin{eqnarray}
 {\rm BR_{tot}}(h_1(1380)  K^{*0})    &=&(14.5\pm4.0)\times 10^{-6},\nonumber\\
 {\rm BR_\parallel}(h_1(1380)  K^{*0})&=&(3.2\pm 1.5)\times 10^{-6},\nonumber\\
 {\rm BR_\perp}(h_1(1380)  K^{*0})&=&(2.0\pm 1.0)\times 10^{-6}
 \end{eqnarray}
in the NP scenario 1, and
 \begin{eqnarray}
 {\rm BR_{tot}}(h_1(1380)  K^{*0})    &=&(8.5\pm 2.0)\times 10^{-6},\nonumber\\
 {\rm BR_\parallel}(h_1(1380)  K^{*0})&=&(2.0\pm 0.5)\times 10^{-6},\nonumber\\
 {\rm BR_\perp}(h_1(1380)  K^{*0})&=&(1.8\pm 0.5)\times 10^{-6}
 \end{eqnarray}
 in the NP scenario 2. The detailed results and discussions can be found in
 Sec.~\ref{sec:results}. We discuss possible NP effects for $\rho K^*$
 modes in Sec.~\ref{sec:conclusion}.

\section{$B\to h_1(1380) K^*$ Standard Model decay
amplitudes}\label{sec:h1ksamp} Within the framework of QCD
factorization, the SM effective Hamiltonian matrix elements are
written in the form of
\begin{equation}\label{fac}
   \langle h_1 \overline K^* |{\cal H}_{\rm eff}|\overline B\rangle
  \! =\! \frac{G_F}{\sqrt2}\!\! \sum_{p=u,c} \! \lambda_p\,
\!   \langle h_1 \overline K^* |\!{\cal T_A}^{h,p}\!+\!{\cal
T_B}^{h,p}\!|\overline B\rangle \,,
\end{equation}
where $\lambda_p\equiv V_{pb}V_{ps}^*$, and the superscript $h $
denotes the final state meson helicity.  ${\cal T_A}$ accounts for
the topologies of the form-factor and spectator scattering, while
${\cal T_B}$ contains annihilation topology amplitudes.

\subsection{Two-parton distribution amplitudes of $1^1P_1$ axial
vector mesons} \label{subsec:DA}

We consider ${\overline B}\to V A$ processes where the $1^1P_1$
axial meson $A$, which is made of $q_1$ and $\bar q_2$, is emitted
from the weak decay vertex in the factorization
amplitudes.~\footnote{If the $^1P_1$ particle is made of $\bar q
q$, then its charge conjugate $C$ is $-1$, i.e., its quantum
number is $J^{PC}=1^{+-}$.}  In the naive factorization,
${\overline B}\to V A$ processes are highly suppressed since
G-parity does not match between the $A$ meson and the axial vector
current $\bar q_1\gamma^\mu\gamma_5 q_2$ in the SU(3) limit. In
the QCD factorization, the QCD radiative corrections can turn the
local operators $\bar q_1\gamma^\mu(1\mp\gamma_5) q_2$ into a
series of nonlocal operators as
 \begin{eqnarray}\label{eq:nonlocal}
  &&  \langle A(P',\epsilon)|\bar q_{1\,\alpha}(y) \, q_{2\,\delta}(x)|0\rangle
= -\frac{i}{4} \, \int_0^1 du \,  e^{i (u \, p'\cdot y +
    \bar u p' \cdot x)}
\nonumber\\
  && \quad \times\,\Bigg\{ f_A m_A \left(
    \not\! p^\prime \gamma_5 \, \frac{\epsilon^*\cdot z}{p'\cdot z} \,
    \Phi_\parallel(u) +  \not\! \epsilon^*_\perp \gamma_5 \,
    g_\perp^{(v)}(u) +  \epsilon_{\mu\nu\rho\sigma} \,
    \epsilon^*{}^\mu  p^{\prime\rho} z^\sigma \, \gamma^\mu
    \, \frac{g_\perp^{(a)}(u)}{4}\right)
\nonumber \\
  && \quad
  - \,f^{\perp}_A \Bigg(\not\!\epsilon_\perp^* \not\! p^\prime \gamma_5\,
  \Phi_\perp(u)
- i \,
   \frac{m_A^2\,\epsilon\cdot z}{(p'\cdot z)^2}
  \, \sigma_{\mu\nu}\gamma_5 \,p^{\prime\mu} z^\nu
  \, h_\parallel^{(t)}(u) - i \, m_A^2 \, \epsilon^*\cdot z \,
  \frac{h_\parallel^{(s)}(u)}{2} \Bigg)
 \Bigg\}_{\delta\alpha}\!\!,
 \end{eqnarray}
where the chiral-even LCDAs are given by
\begin{eqnarray}
  &&\langle A(P',\epsilon)|\bar q_1(y) \gamma_\mu \gamma_5 q_2(x)|0\rangle
  = if_A m_A \, \int_0^1
      du \,  e^{i (u \, p'\cdot y +
    \bar u p' \cdot x)}
   \left\{p^\prime_\mu \,
    \frac{\epsilon^*\cdot z}{p'\cdot z} \, \Phi_\parallel(u)
         +\epsilon_\perp^*{}_\mu \, g_\perp^{(v)}(u)
         \right\}, \nonumber\\ \label{eq:evendef1} \\
  &&\langle A(P',\epsilon)|\bar q_1(y) \gamma_\mu
  q_2(x)|0\rangle
  = - i f_A m_A \,\epsilon_{\mu\nu\rho\sigma} \,
      \epsilon^*{}^\nu p^{\prime\rho} z^\sigma \,
    \int_0^1 du \,  e^{i (u \, p'\cdot y +
    \bar u p' \cdot x)} \,
       \frac{g_\perp^{(a)}(u)}{4}, \label{eq:evendef2}
\end{eqnarray}
with the matrix elements involving an odd number of $\gamma$
matrices and $\bar u\equiv 1-u$, and the chiral-odd LCDAs are
given by
\begin{eqnarray}
  &&\langle A(P',\epsilon)|\bar q_1(y) \sigma_{\mu\nu}\gamma_5 q_2(x)
            |0\rangle
  =  f_A^{\perp} \,\int_0^1 du \, e^{i (u \, p'\cdot y +
    \bar u p' \cdot x)} \,
\Bigg\{(\epsilon^*_\perp{}_\mu p_{\nu}^\prime -
  \epsilon_\perp^*{}_\nu  p^\prime_{\mu}) \,
  \Phi_\perp(u),\nonumber\\
&& \hspace*{+5cm}
  + \,\frac{m_A^2\,\epsilon^*\cdot z}{(p'\cdot z)^2} \,
   (p^\prime_\mu z_\nu -
    p^\prime_\nu  z_\mu) \, h_\parallel^{(t)}(u)
\Bigg\},\label{eq:odddef1}\\
&&\langle A(P',\epsilon)|\bar q_1(y) \gamma_5 q_2(x)
            |0\rangle
  =  f_A^{\perp} m_{A}^2 \epsilon^*\cdot z\,\int_0^1 du \, e^{i (u \, p'\cdot y +
    \bar u p' \cdot x)}  \, \frac{h_\parallel^{(s)}(u)}{2},\label{eq:odddef2}
\end{eqnarray}
with the matrix elements containing an even number of $\gamma$
matrices. Here, throughout the present discussion, we define
$z=y-x$ with $z^2=0$, and introduce the light-like vector
$p'_\mu=P'_\mu-m_A^2 z_\mu/(2 P'\cdot z)$ with the meson's
momentum satisfying ${P'}^2=m_A^2$. Moreover, the meson
polarization vector $\epsilon_\mu$ has been decomposed into
longitudinal and transverse {\it projections} defined as
\begin{eqnarray}\label{eq:polprojectiors}
 && \epsilon^*_\parallel{}_\mu \equiv
     \frac{\epsilon^* \cdot z}{P'\cdot z} \left(
      P'_\mu-\frac{m_A^2}{P'\cdot z} \,z_\mu\right), \qquad
 \epsilon^*_\perp{}_\mu
        = \epsilon^*_\mu -\epsilon^*_\parallel{}_\mu\,,
\end{eqnarray}
respectively. The LCDAs $\Phi_\parallel, \Phi_\perp$ are of
twist-2, and $g_\perp^{(v)}, g_\perp^{(a)}, h_\perp^{(t)},
h_\parallel^{(s)}$ of twist-3. Due to G-parity, $\Phi_\parallel,
g_\perp^{(v)}$ and $g_\perp^{(a)}$ are antisymmetric with the
replacement $u\to 1-u$, whereas $\Phi_\perp, h_\parallel^{(t)}$
and $h_\parallel^{(s)}$ are symmetric in the SU(3) limit. We
restrict ourselves to two-parton LCDAs with twist-3 accuracy.

To perform the calculation in the momentum space, we first
represent Eq.~(\ref{eq:nonlocal}) in terms of $z$-independent
variables, $P'$ and $\epsilon^*$. For simplicity, we introduce two
light-like vectors $n_-^\mu\equiv (1,0,0,-1), n_+^\mu\equiv
(1,0,0,1)$. If neglecting the meson mass squared, we have
$p^{\prime\, \mu} = E n_-^\mu$ where $E$ is the energy of the $A$
meson in the $B$ rest frame. Choosing the momentum of the quark
$q_1$ in the $A$ meson as
\begin{eqnarray}
k_1^\mu = u E n_-^\mu +k_\perp^\mu + \frac{k_\perp^2}{4
uE}n_+^\mu\,,
\end{eqnarray}
we apply the following substitution in the calculation
\begin{equation}
z^\mu \to -i \frac{\partial}{\partial k_{1\, \mu}}\simeq -i \Bigg(
\frac{n_+^\mu}{2E}\frac{\partial}{\partial u} +
\frac{\partial}{\partial k_{\perp\, \mu}}\Bigg)\,,
\end{equation}
where the term of order $k_\perp^2$ is omitted. Note that all the
components of the coordinate $z$ should be taken into account in
the calculation before the collinear approximation is applied.
Then, the light-cone projection operator of an $A$ meson in the
momentum space reads
\begin{equation}
  M_{\delta\alpha}^A =  M_{\delta\alpha}^A{}_\parallel +
   M_{\delta\alpha}^A{}_\perp\,,
\label{rhomeson2}
\end{equation}
with the longitudinal part
 \begin{eqnarray}
M^A_\parallel &=& -i\frac{f_A}{4} \, \frac{m_A(\epsilon^*\cdot
  n_+)}{2}
 \not\! n_- \gamma_5\,\Phi_\parallel(u)
-\frac{if_A^\perp m_A}{4}  \,\frac{m_A(\epsilon^*\cdot n_+)}{2E}
 \, \Bigg\{\frac{i}{2}\,\sigma_{\mu\nu}\gamma_5 \,  n_-^\mu  n_+^\nu \,
 h_\parallel^{(t)}(u)
\nonumber\\
&& \hspace*{-0.0cm} + \,i E\int_0^u dv \,(\Phi_\perp(v) -
h_\parallel^{(t)}(v)) \
     \sigma_{\mu\nu} \gamma_5 n_-^\mu
     \, \frac{\partial}{\partial k_\perp{}_\nu}
  - \gamma_5\frac{h_\parallel'{}^{(s)}(u)}{2}\Bigg\}\, \Bigg|_{k=u p'}\,,
\end{eqnarray}
 and the transverse part
 \begin{eqnarray}
M^A_\perp &=& i\frac{f^{\perp}_A}{4} E\not\!
\epsilon^*_\perp\not\! n_- \gamma_5 \,
   \Phi_\perp(u)\nonumber\\
&&  -i \frac{f_Am_A}{4} \,\Bigg\{\not\! \epsilon^*_\perp\gamma_5
\, g_\perp^{(v)}(u) -  \, E\int_0^u dv\, (\Phi_\parallel(v) -
g_\perp^{(v)}(v))
       \not\! n_-\gamma_5 \, \epsilon^*_{\perp\mu} \,\frac{\partial}{\partial
         k_{\perp\mu}}
\cr && + \,i \epsilon_{\mu\nu\rho\sigma} \,
        \epsilon_\perp^{*\nu} \, \gamma^\mu n_-^\rho
         \left[n_+^\sigma \,{1\over 8}\frac{dg_\perp^{(a)}(u)}{du}-
          E\,\frac{g_\perp^{(a)}(u)}{4} \, \frac{\partial}{\partial
         k_\perp{}_\sigma}\right]
 \Bigg\}
 \, \Bigg|_{k=up'},
\end{eqnarray}
where the transverse polarization vectors of the axial vector
meson are
\begin{equation}\label{app:polvector}
\epsilon_\perp^\mu \equiv \epsilon^\mu - \frac{\epsilon\cdot
n_+}{2}\,n_-^\mu- \frac{\epsilon\cdot n_-}{2}\,n_+^\mu \,,
\end{equation}
which is, instead of that in Eq.~(\ref{eq:polprojectiors}),
independent of the coordinate. In the present study, we choose the
coordinate systems in the Jackson convention~\cite{Das:2004hq},
which is adopted by BaBar and Belle measurements. In other words,
in the $\overline B$ rest frame, if the $z$ axis of the coordinate
system is along the the direction of the flight of the $V$ meson,
we can have
 \begin{eqnarray}
 &&\epsilon_{V}^{\mu}(0)=(p_c, 0, 0, E_V)/m_V,\ \ \ \
   \epsilon_{A}^{\mu}(0)=(p_c, 0, 0, -E_{A})/m_{A}, \nonumber\\
 &&\epsilon_{V}^{\mu}(\pm 1)=\frac{1}{\sqrt{2}}(0, \mp 1, -i, 0),
 \ \ \ \
   \epsilon_{A}^{\mu}(\pm 1)=\frac{1}{\sqrt{2}}(0, \mp 1, +i,
   0),
 \end{eqnarray}
where $p_c$ is the center mass momentum of the final state meson.
In the large energy limit, we have $\epsilon^*_A(\lambda)\cdot n_+
=2E_A/m_A\, \delta_{\lambda,0}$ and $\epsilon^*_A(\lambda)\cdot
n_- =0$. Note that if the coordinate systems are chosen in the
Jacob-Wick convention~\cite{Das:2004hq}, the transverse
polarization vectors of the $A$ meson become
$\epsilon_{A}^{\mu}(\pm 1)=(0, \pm 1, -i, 0)/\sqrt{2}$. In
general, the QCD factorization amplitudes can be reduced to the
form of $\int_0^1 du \, {\rm tr} (M^A\dots)$.

In the following, we will give a brief discussion for LCDAs of $V$
and $A$ mesons. The detailed information for LCDAs of the vector
mesons can be found in \cite{Beneke:2000wa,Das:2004hq}. The
asymptotic twist-2 distribution amplitudes are
 \begin{eqnarray}\label{eq:lcdas}
 \Phi_\parallel^{V}(u)=\Phi_\perp^{V}(u) =\Phi_\perp^{A}(u)=
 6u\bar u\,,
 \end{eqnarray}
 but $\Phi_\parallel^{A}(u)=0$ in SU(3)
due to G-parity. $\Phi_\parallel^{A}(u)$ can be expanded in
Gegenbauer polynomials with only odd terms:
\begin{eqnarray}\label{eq:phiparallel}
\Phi_\parallel^{A}(u)= 6u\bar u \bigg[\sum_{i=1,3,5,\dots}
a_i^{A,\parallel} C_i^{3/2}(2u-1)\bigg]\,,
\end{eqnarray}
where we have neglected the even terms due to possible $m_{q_1}
\not = m_{q_2}$. Note that since the product
$f_{A}a_1^{A,\parallel}$ always appears together, we simply take
$f_{A}=f_{A}^\perp$ in the present study, while
$a_1^{A,\parallel}$ is determined in Ref.~\cite{kcymoments}. If
neglecting the three-parton distributions and terms proportional
to the light quark masses, the  twist-3 distribution amplitudes
for both $V$ and $A$ mesons can be related to the twist-2 ones by
Wandzura-Wilczek relations~\cite{Ball:1998sk,kcymoments}:
\begin{eqnarray}\label{eq:ww}
 && h_\parallel^{\prime(s)}(v)= -2\Bigg[ \int_0^v
\frac{\Phi_\perp(u)}{\bar u}du -\int_v^1
\frac{\Phi_\perp(u)}{u}du \Bigg], \nonumber\\
 && h_\parallel^{\prime(t)}(v)= (2u-1)\Bigg[ \int_0^v
\frac{\Phi_\perp(u)}{\bar u}du -\int_v^1
\frac{\Phi_\perp(u)}{u}du \Bigg], \nonumber\\
 &&\int_0^v du \big( \Phi_\perp (u) -h^{(t)}_\parallel
(u))= v\bar v\Bigg[ \int_0^v \frac{\Phi_\perp(u)}{\bar u}du
-\int_v^1
\frac{\Phi_\perp(u)}{u}du\Bigg], \nonumber\\
&&\int_0^v du \big( \Phi_\parallel (u) -g^{(v)}_\perp (u))=
 \frac{1}{2}\Bigg[ \bar v\int_0^v \frac{\Phi_\parallel(u)}{\bar u}du
 -v \int_v^1 \frac{\Phi_\parallel(u)}{u}du\Bigg], \nonumber\\
 && g_\perp^{(a)}(v)= 2\Bigg[ \bar v\int_0^v
\frac{\Phi_\parallel(u)}{\bar u}du + v\int_v^1
\frac{\Phi_\parallel(u)}{u}du \Bigg], \nonumber\\
 && \frac{g_\perp^{\prime(a)}(v)}{4}+g_\perp^{(v)}(v)= \int_v^1
\frac{\Phi_\parallel (u)}{u}du\,,\nonumber\\
&& \frac{g_\perp^{\prime(a)}(v)}{4}-g_\perp^{(v)}(v)=- \int_0^v
\frac{\Phi_\parallel (u)}{\bar u}du\,.
\end{eqnarray}

\subsection{$\overline B \to h_1(1380) \overline K^*$ amplitudes
with topologies of the form-factor and spectator scattering}

${\cal T_A}^{h,p}$ describes contributions from naive
factorization, vertex corrections, penguin contractions and
spectator scattering. However, for $\overline B \to h_1(1380)
\overline K^*$ processes, the naive factorization amplitudes are
forbidden due to the mismatch of the G-parity between the
$h_1(1380)$ meson and the local axial-vector current $\bar
s\gamma^\mu\gamma_5 s$. The resultant amplitude reads
\begin{eqnarray}
&& \frac{G_F}{\sqrt2}\!\! \sum_{p=u,c} \! \lambda_p\, \!   \langle
h_1 \overline K^{*} |\!{\cal T_A}^{h,p} |\overline B\rangle
\nonumber\\
 && = \frac{G_F}{\sqrt 2} \left(- V_{tb} V_{ts}^*\right)
\left[a_3^h + a_4^h - a_5^h + r_\chi^{h_1} a_6^h -
\frac{1}{2}(-a_7^h + r_\chi^{h_1} a_8^h + a_9^h + a_{10}^h)\right]
X_h^{({\overline B} {\overline K^{*}}, h_1)} ,
\end{eqnarray}
where~\footnote{There is a relative sign difference between
$X^{({\overline B} {\overline K^{*}}, h_1)}_0$ and
 $X^{({\overline B} {\overline K^{*}}, h_1)}_\pm$ due to the adoption of the
Jackson convention for the coordinate systems.}
 \begin{eqnarray}
&& X^{({\overline B} {\overline K^{*}}, h_1)}_0 \nonumber\\
&&\ \ \ =
-\frac{f_{h_1}}{2m_{K^*}}\Bigg[(m_B^2-m_{K^*}^2-m_{h_1}^2)(m_B+m_{K^*})A^{B{K^*}}_1(m_{h_1}^2)
 -{4m_B^2p_c^2\over m_B+m_{K^*} }\,A^{B{K^*}}_2(m_{h_1}^2)\Bigg], \nonumber \\
&& X^{({\overline B} {\overline K^{*}}, h_1)}_\pm = f_{h_1}
m_{h_1} \Bigg[ (m_B+m_{K^*})A^{B{K^*}}_1(m_{h_1}^2)\mp
{2m_Bp_c\over m_B+m_{K^*}}\,
  V^{B{K^*}}(m_{h_1}^2) \Bigg],\end{eqnarray}
with $q = p_B - p_{K^*}\equiv p_{h_1}$, $p_c$ being the center
mass momentum of the final state mesons in the $\overline B$ rest
frame and
 \begin{eqnarray}
 r_\chi^{h_1} = \frac{2
 m_{h_1}}{m_b(\mu)}\frac{f_{h_1}^\perp(\mu)}{f_{h_1}}.
 \end{eqnarray}
 Here the form factors are defined as
\begin{eqnarray}
\langle{\overline K^{*}}(p_{K^*},
 \epsilon_{K^*})|V_\mu|{\overline B} (p_B)\rangle
 &=& \frac{2 }{m_B + m_{K^*}} \epsilon_{\mu\nu\alpha\beta} \epsilon_{K^*}^{*\nu}
 p_B^\alpha p_{K^*}^{\beta} V(q^2),
\nonumber \\
 \langle{\overline K^{*}}(p_{K^*},
 \epsilon_{K^*})|A_\mu|{\overline B} (p_B)\rangle
 &=& i \left[(m_B + m_{K^*}) \epsilon_{K^*\mu}^{*} A_1(q^2)
 - (\epsilon_{K^*}^* \cdot p_B)
(p_B + p_{K^*})_\mu \frac{ A_2(q^2)}{m_B + m_{K^*}}\right]
\nonumber \\
&& - 2 i m_{K^*} \frac{\epsilon_{K^*}^*\cdot p_B}{q^2} q^\mu
\left[A_3(q^2) - A_0(q^2)\right],
 \end{eqnarray}
where   $A_3(0) = A_0(0)$ and
 \begin{eqnarray}
 A_3(q^2)= \frac{m_B + m_{K^*}}{2 m_{K^*}} A_1(q^2) - \frac{m_B -
m_{K^*}}{2 m_{K^*}} A_2(q^2).
 \end{eqnarray}
 In general, for ${\overline B}\to V A$ processes with the $A$
meson emitted from the weak decay vertex, $a_i^h$'s are given by
 \begin{eqnarray} \label{eq:ai}
 a_1^h &=& {\alpha_s\over 4\pi}\,
 {C_F\over N_c}c_2\,(f_{I}^{h,0}+f_{II}^{h,0}),
\nonumber \\
 a_2^h &=& {\alpha_s\over 4\pi}\,{C_F\over
N_c}c_1\,(f_{I}^{h,0}+f_{II}^{h,0}), \nonumber
 \\
 a_3^h &=& {\alpha_s\over 4\pi}\,{C_F\over N_c} c_4\,(f_{I}^{h,0}+f_{II}^{h,0}),
  \nonumber \\
 a_4^h &=& {\alpha_s\over 4\pi}\,{C_F\over N_c}
 \Bigg\{c_3 (f_{I}^{h,0}+f_{II}^{h,0})+G^h(s_s)+G^h(s_b)\big]-c_1
 \left({\lambda_u\over \lambda_t}G^h(s_u)+{\lambda_c\over\lambda_t}G^h(s_c)\right)
\nonumber \\
 && +(c_4+c_6) \sum_{i=u}^b G^h(s_i) +{3\over 2}
(c_8+c_{10})\sum_{i=u}^b e_i G^h(s_i)
 +{3\over 2}c_9\big[e_{q'} G^h(s_{q'})-{1\over 3}G^h(s_b)\big]+c_g G^h_g\Bigg\},
  \nonumber \\
 a_5^h &=& -{\alpha_s\over 4\pi}\,{C_F\over N_c} c_6( f_{I}^{h,1} + f_{II}^{h,1}),\nonumber\\
 a_6^h &=&  {\alpha_s\over 4\pi}\,{C_F\over N_c}  \Bigg\{ c_5 \tilde f_{I}^{h}
  - c_1 \left({\lambda_u\over \lambda_t}\hat G^h(s_u)
  + {\lambda_c\over\lambda_t}\hat G^h(s_c)\right)
  +c_3\big[\hat G^h(s_s)+ \hat G^h(s_b)\big]\nonumber\\
 && + (c_4+c_6) \sum_{i=u}^b \hat G^h(s_i) \Bigg\},\nonumber\\
 a_7^h &=& -{\alpha_s\over 4\pi}\,{C_F\over N_c} c_8( f_{I}^{h,1}+
 f_{II}^{h,1}), \nonumber \\
 a_8^h &=&   {\alpha_s\over 4\pi}\,{C_F\over N_c} c_7 \tilde f_{I}^{h}
 -{\alpha\over 9\pi}\,\hat C^h_e, \nonumber\\
 a_9^h &=& {\alpha_s\over 4\pi}\,{C_F\over N_c}
 c_{10}\,(f_{I}^{h,0}+f_{II}^{h,0}) ,  \nonumber \\
 a_{10}^h &=& {\alpha_s\over 4\pi}\,{C_F\over N_c}
c_9\,(f_{I}^{h,0}+f_{II}^{h,0})-{\alpha\over 9\pi}\,C^h_e,
 \end{eqnarray}
  where $c_i$ are the Wilson coefficients,
$C_F=(N_c^2-1)/(2N_c)$, $s_i=m_i^2/m_b^2$, and $\lambda_q =V_{qb}
V_{qq'}^*$, with $q'=d,s$. In Eq.~(\ref{eq:ai}), the vertex
corrections are given by
\begin{eqnarray}
 f_I^{0,i} &=& \int^1_0 dx\,\Phi_\parallel^{A}(x)
 \bar g^i(x) \,, \nonumber \\
f_I^{\pm,i} &=& \int^1_0 dx\,\Bigg( g_\perp^{A (v)}(x)\pm (-1)^i
{1\over 4} \frac{d g_\perp^{A (a)}(x)}{dx} \Bigg)
 \bar g^i(x) \,,\nonumber\\
  \tilde f_I^{0} &=&  \int^1_0 dx\, \frac{-h_\parallel^{'(s)A}(x)}{2}
  \Big[ 2{\rm Li}_2(x) -\ln^2x -(1+2i\pi)\ln x -
 (x \leftrightarrow 1-x)\Big]\,, \nonumber \\
  \tilde f_I^{\pm} &=& 0\,, \label{fI}
 \end{eqnarray}
with
 \begin{eqnarray}
 \bar g^0(x)&=& 3\left( \frac{1-2x}{1-x} \right)\ln x\nonumber\\
  && + \Big[ 2{\rm Li}_2(x) -\ln^2x +\frac{2\ln x}{1-x} -(3+2i\pi)\ln x -
 (x \leftrightarrow 1-x)\Big]\,, \nonumber\\
 \bar g^1(x)&=&\bar g^0(1-x) \,.
 \end{eqnarray}
 Here $f_{II}^{h,i}$, arising from the hard
spectator interactions with a hard gluon exchange between the
emitted $h_1(1380)$ meson and the spectator quark of the
$\overline B$ meson, have the expressions:
 \begin{eqnarray}
  f_{II}^{0,i} &=& {4\pi^2\over N_c}\,
  {f_B f_{A} f_{V} \over 2 X^{({\overline B}^0 {V}, A)}_0}
  \int^1_0 d\rho\, {\Phi^B_1(\rho)\over \drho}\int^1_0 d v\,
 \Phi^{V}_\parallel(v)   \nonumber\\
 &&\ \times \int^1_0 d u \,
 \Bigg[ (-1)^i \Bigg(\frac{1}{u \bar v}-\frac{1}{\bar u \bar v} \Bigg)
 \Bigg(\Phi^{A}_\parallel(u) - r_\chi^{A} \frac{h^{\prime(s)A}_\parallel(u)}{2}\Bigg) \nonumber\\
 &&\ \ \ \ \ -
 \Bigg(\frac{1}{u \bar v}+\frac{1}{\bar u \bar v}\Bigg)
 \Bigg(\Phi^{A}_\parallel(u) + r_\chi^{h_1} \frac{h^{\prime(s)A}_\parallel(u)}{2}\Bigg)\Bigg], \nonumber \\
 f_{II}^{\pm,i} &=&  - {4\pi^2\over N_c}\, {2 f_B f_{A}
 f^{\perp}_{V} m_{A} \over m_B X^{({\overline B}^0 {V}, A)}_\pm}
 (1\mp 1)\int^1_0 d\rho\, {\Phi^B_1(\rho)\over \drho}\int^1_0 dv\,
 {\Phi^{V}_\perp(v)\over \bar v^2}\int^1_0 d u\, \nonumber \\
  &&\times \Bigg[\Bigg( g_\perp^{A (u)}(u)
  - (-1)^i {1\over 4}\frac{d g_\perp^{A (a)}(u)}{d u}\Bigg)
   +
 \Bigg( \frac{1}{u} - \frac{1}{\bar u} \Bigg)
 \int_0^u dx (\Phi_\parallel^{A}(x) -g_\perp^{A (v)} (x))
 \Bigg] \nonumber\\
 && - {4\pi^2\over N_c}\,{f_B f_{A} f_{V} m_{A} m_{V}\over m_B^2
  X^{({\overline B}^0 V, A)}_\pm}\int^1_0 d\rho\,
  {\Phi^B_1(\rho)\over\drho}\int^1_0 dv\,\Bigg(g^{V(v)}_\perp(v)
 \pm {1\over 4} {d g^{V(a)}_\perp(v)\over d v} \Bigg)\nonumber\\
 && \ \ \ \times \int^1_0 du
   \Bigg\{ \Bigg[ (-1)^i \Bigg( \frac{u +\bar v}{u \bar v^2}
  - \frac{\bar u +\bar v}{\bar u \bar v^2} \Bigg)
  - \Bigg( \frac{u +\bar v}{u \bar v^2}
  +\frac{\bar u +\bar v}{\bar u \bar v^2} \Bigg)\Bigg]
\nonumber\\
 &&\ \ \ \ \times
 \Bigg(g_\perp^{A (v)}(u) \pm (-1)^i {1\over 4}{d g_\perp^{A (a)}(u) \over du}
  \Bigg) \nonumber\\
  &&\ \ \ -  \Bigg[ (-1)^i \Bigg( \frac{1}{u \bar v^2}
  + \frac{1}{\bar u \bar v^2} \Bigg)
  + \Bigg( \frac{1}{u \bar v^2}
  - \frac{1}{\bar u \bar v^2} \Bigg) \Bigg]
 \int_0^u dx (\Phi_\parallel^{A}(x) -g_\perp^{A (v)} (x))\Bigg\},
 \label{fII2}
\end{eqnarray}
with $\Phi^B_1(\rho)$ being one of the two light-cone distribution
amplitudes of the $\overline B$ meson~\cite{CY,Beneke:2000wa}.
$G^h, \hat G^h, C_e^h$ and $\hat C_e^h$, originating from QCD and
electroweak contractions, respectively, are given by
 \begin{eqnarray}
 G^0(s) &=& 4\int^1_0 du\,\Phi^{A}_\|(u)\int^1_0 dx\,x\bar x
  \ln[s-\bar u x\bar x-i\epsilon], \nonumber \\
 G^\pm(s) &=& 2\int^1_0 du\,
  \Bigg( g^{A (v)}_\perp(u) \pm {1\over 4}{d g_\perp^{A (a)}(u) \over du }\Bigg)
  \int^1_0 dx\,x\bar x\ln[s-\bar u x\bar x-i\epsilon]\,,\nonumber\\
 \hat G^0(s) &=& 4\int^1_0 du\,\frac{-h^{\prime(s)A}_\parallel(u)}{2}\int^1_0 dx\,x\bar x
  \ln[s-\bar u x\bar x-i\epsilon], \nonumber \\
 \hat G^\pm(s) &=& 0\,,\nonumber\\
 C_e^h (s) &=& \left({\lambda_u\over \lambda_t}G^h(s_u)
 +{\lambda_c\over \lambda_t}G^h(s_c)\right) \left(c_2+{c_1\over N_c}\right)\,,
 \end{eqnarray}
and $\hat C_e^h$ can be obtained from $C_e^h$ with the replacement
$G^h \to \hat G^h$, where small electroweak corrections from
$c_{7-10}$ are neglected in $C_e^h$ and $\hat C_e^h$. The dipole
operator $O_g$ gives
  \begin{eqnarray}
&&  G^0_g = -2\int^1_0 du\,{\Phi^{A}_\|(u)\over \bar u}\,, \nonumber\\
&&  G^\pm_g =\int^1_0 {du\over \bar u}\,\Bigg[ \int_0^u\,
\Big(\Phi_\|^{A}(v)-g^{A (v)}_\perp(v)\Big)dv
  -\bar u g_\perp^{A (v)}(u) \mp {{\bar u} \over 4}{d g_\perp^{A (a)}(u) \over du} +
{g_\perp^{A (a)}(u) \over 4}\Bigg]. \hspace{0.7cm} \label{eq:cg}
 \end{eqnarray}
Using Eq.~(\ref{eq:ww}), $G_g^\pm$ can be further reduced to
\begin{eqnarray}
G_g^+ &=& \int_0^1 du \Bigg( \int_0^u
 \frac{\Phi^{A}_\parallel(v)}{\bar v}dv
 - \int_u^1 \frac{\Phi^{A}_\parallel(v)}{v}dv \Bigg) =0, \nonumber\\
G_g^- &=& 0,
\end{eqnarray}
where we have taken the approximation
$\Phi_\parallel^{A}(u)=6u\bar u  a_1^{A,\parallel}
C_1^{3/2}(2u-1)$. Obviously, considering only two-parton
distribution amplitudes, the dipole operator does not contribute
to transverse amplitudes at ${\cal O}(\alpha_s)$. The result is
consistent with the fact that in large $m_b$ limit the transverse
amplitudes are suppressed since the outgoing $s$ and $\bar s$
arising from $s-\bar s-n\ gluons$ couplings have opposite
helicities. Note that the linear infrared divergence, originating
from twist-3$\times$twist-3 final-state LCDAs~\footnote{We have
checked that the linear divergence is not cancelled by
twist-4$\times$twist-2 ones.}, is present in $f_{II}^\pm$ and it
may exist a mechanism in analogy to the heavy-light transition
form factors where the linear divergences are consistently
absorbed into the form factors~\cite{Beneke:2000wa}. However we
will introduce a infrared cutoff, $\Lambda_{\rm QCD}/m_b$, to
regulate the linear divergence. The numerical results are very
insensitive to the cutoff, and, moreover, the transverse
contributions are already suppressed. On the other hand, we shall
parameterize the logarithmic divergence, appearing in $f_{II}^h$,
as
 \begin{eqnarray}
 \int_0^1 \frac{dx}{\bar x}\to X_H^h=\ln
 \left(\frac{m_B}{\Lambda_h}\right)(1+\rho_H^h
 e^{i\phi_H^h})\,,
 \end{eqnarray}
with $\rho_H^h \leq 1$ and $\Lambda_h\approx 0.5$~GeV.

\subsection{$\overline B \to h_1(1380) \overline K^*$ amplitudes
with topologies of annihilation}

We shall see  that $\overline B\to h_1(1380) \overline K^*$
helicity amplitudes in the SM may be governed by the annihilation
topologies. The weak annihilation contributions to $\overline
B^0\to h_1(1380) \overline K^{*0}$ read
\begin{eqnarray}\label{eq:h1ksann}
\frac{G_F}{\sqrt2}\!\! \sum_{p=u,c} \! \lambda_p\, \!   \langle
h_1 \overline K^{*0} |\!{\cal T_B}^{h,p} |\overline B^0\rangle
&\simeq& \frac{G_F}{\sqrt{2}}\sum_{p=u,c} V_{pb}V_{ps}^*
 \Bigg\{ f_Bf_{K^*} f_{h_1} \bigg[b_3^h
 - \frac{1}{2}b_{\rm 3,EW}^h\bigg] \Bigg\},
\end{eqnarray}
where
\begin{eqnarray}\label{eq:bi}
 b_3^h &=& \frac{C_F}{N_c^2} \Big[ c_3 A_1^{i,h}
  \!  +c_5 A_3^{i,h} \!+ (c_5 + N_c c_6 ) A_3^{f,h}
  \Big]\,,\nonumber\\
b^h_{3\,EW}  &=& \frac{C_F}{N_c^2} \Big[ c_9 A_1^{i,h}
  \!  +c_7 A_3^{i,h} \!+ (c_7 + N_c c_8 ) A_3^{f,h} \Big]\,.
 \end{eqnarray}
For $B^-\to h_1(1380) K^{*-}$ an additional term $\delta_{pu}
b_2^h$ is needed in the square bracket in Eq.~(\ref{eq:h1ksann})
with $b_2^h = c_2 A_1^{i,h} C_F/N_c^2$. The annihilation
amplitudes originating from operators $(\bar q_1 b)_{V-A}
(\bar{q}_2 q_3 )_{V-A} $ and $-2(\bar q_1 b)_{S-P} (\bar{q}_2 q_3
)_{S+P} $ are denoted as $A_{1}^{i,f\, (h)}$ and
$A_{3}^{i,f\,(h)}$, respectively, where the superscript $i \,(f)$
indicates gluon emission from the initial (final) state quarks in
the weak vertex. In Eq.~(\ref{eq:h1ksann}), we will neglect the
terms proportional to $A_{1(3)}^{i,h}$ due partially to
$c_{3,5,9}\ll (c_5+N_c c_6)$ or CKM suppression (for $B^-$).
Although $A_1^{i,-}$ contains a linear divergence arising from the
twist-3$\times$twist-3 final state distribution amplitudes, it was
argued in Ref.~\cite{Kagan} that the divergence should be
cancelled by the twist-4$\times$twist-2 ones. Moreover,
$A_1^{i,-}$ is still relatively small compared to
$A_3^{f,-}$~\cite{Kagan}. $A_3^{f,h}$, for which one of the final
state mesons arises with the twist-3 distribution amplitude, while
the other is of twist-2, are given by
\begin{eqnarray}
A_3^{f,\,0}(h_1 \overline K^*) &=& \pi\alpha_s \int_0^1\! du
\,dv\, \Bigg[ \frac{2 m_{h_1} f_{h_1}^\perp}
 {m_b f_{h_1} }\,\Phi^{K^*}_{\parallel}(v)\,
\bigg( \frac{ h_\parallel^{h_1 \prime(s)}(u)}{2}- \frac{v}{\bar u}
\int_0^u dx \big( \Phi^{h_1}_\perp (x) -h^{h_1(t)}_\parallel
(x))\bigg)
\frac{2}{v^2 \bar u} \nonumber\\
&& + \frac{2 m_{K^*} f_{K^*}^\perp}
 {m_b  f_{K^*}}\,\Phi^{h_1}_{\parallel}(u)\,
\bigg( \frac{h_\parallel^{K^*\prime(s)}(v)}{2} - \frac{\bar u}{v}
\int_0^v dx \big( \Phi^{K^*}_\perp (x) -h^{K^*(t)}_\parallel
(x))\bigg)
\frac{2}{\bar u^2 v}\Bigg]\,,\label{A3f0} \\
 A_3^{f,\,-}(h_1 \overline K^*)  &=& \!
\pi\alpha_s \int_0^1\! du \,dv\, \Bigg[  \frac{2 m_{h_1}
f_{K^*}^\perp}
 {m_b f_{K^*} }\,\Phi^{K^*}_{\perp}(v)\,
 \Bigg( g_\perp^{h_1 (v)}(u)- \frac{g_\perp^{h_1\prime(a)}(u)}{4} \Bigg)\,
\frac{2}{v^2\bar u }\nonumber\\
 &&\ \ \ \ \ \ \ \ \ \ \ \ \ \ \ - \frac{2 m_{K^*} f_{h_1}^\perp}
 {m_b f_{h_1}}\,\Phi^{h_1}_{\perp}(u)\,
 \Bigg( g_\perp^{K^*(v)}(v) + \frac{g_\perp^{K^*\prime(a)}(v)}{4} \Bigg)\,
\frac{2}{\bar u^2 v}\Bigg]\,,\label{A3fm}
 \end{eqnarray}
where the detailed definitions of the distribution amplitudes of
the $h_1$ meson have been collected in Sec.~\ref{subsec:DA}.

Using the asymptotic distribution amplitudes of
$\Phi_{\parallel,\perp}^{K^*}(u)$ and $\Phi_\perp^{h_1}(u)$, and
the approximation for $\Phi_\parallel^{h_1}(u)=6u\bar u  a_1^{h_1}
C_1^{3/2}(2u-1)$ given in Eqs.~(\ref{eq:lcdas}) and
(\ref{eq:phiparallel}), respectively,
we obtain the annihilation amplitudes
\begin{eqnarray}
 A_3^{f,\,0}(h_1 \overline K^*) & \approx& - 18 \pi\alpha_s
(X_A^0-2) \bigg[\frac{2 m_{h_1} f_{h_1}^\perp} {m_b f_{h_1} }
(2X_A^0-1) -\frac{2 m_{K^*} a_1^{h_1,\parallel}f_{K^*}^\perp}{m_b
f_{K^*}}
(6X_A^0-11)\bigg]\,,\label{eq:A0} \hspace{1cm} \\
A_3^{f,\,-}(h_1 \overline K^*) & \approx& -18 \pi\alpha_s
(X^-_A-1) \Bigg[ \frac{2 m_{K^*} f_{h_1}^\perp} {m_b f_{h_1} } ( 2
X^-_A -3)
 -\frac{2 m_{h_1} a_1^{h_1,\parallel}f_{K^*}^\perp}
 {m_b f_{K^*}}\biggl( 2 X^-_A -\frac{17}{3}\biggl)\Bigg],
 \label{eq:Am}\hspace{1cm}
 \end{eqnarray}
where the logarithmic divergences are parameterized as $X_A^h = (1
\!+\! \rho_A^h \,e^{i \varphi^{h}_A} ) \ln\,( {m_B}/{\Lambda_h})$
with $\rho_A^h \lsim 1$. Compared with $\overline A_0^{SM}$ given
in Eq.~(\ref{eq:smamp}), the annihilation amplitudes thus
contribute to the longitudinal and negative polarized states as
${\cal O}[1/m_b^2 \ln^2 (m_b/\Lambda_h)]$, where we use that $f_B
\sim1/m_b^{-1/2}$~\cite{BBNS2} and $\zeta_\parallel \sim
1/m_b^{-3/2}$~\cite{Charles:1998dr}. We obtain
$f_{h_1}^\perp(1~{\rm GeV})\simeq 0.2$~GeV and $f_{h_1}
a_1^{h_1,\parallel}(1~{\rm GeV}) \approx -0.45$~GeV from the QCD
sum rule calculation~\cite{kcymoments}, where the numerical values
will be listed in Sec.~\ref{sec:results}. In Eq.~(\ref{eq:A0}),
the second term in $A_3^{f,0}$ is numerically dominant such that
annihilation corrections are constructive to the longitudinal
amplitude, while in Eq.~(\ref{eq:Am}), two terms in $A_3^{f,-}$
tend to cancel each other such that annihilation effects are
negligible for transverse fractions. For comparison, we list
$A_3^{f,0 (-)}$ for the $\overline B\to \phi \overline K^*$ decays
as follows:
\begin{eqnarray}
\label{eq:XAA3}
  && A_3^{f,\,0} (\phi \overline K^*) \approx  18i \pi\alpha_s
 \bigg(
 \frac{2 m_{\phi} f_{\phi}^\perp }{m_b f_{\phi}}
  + \frac{2 m_{K^*} f_{K^*}^\perp }{m_b f_{K^*}} \bigg)
(X_A^0-2)(2 X_A^0-1)  \,, \nonumber \\
 && A_3^{f,\,-} (\phi \overline K^*) \approx 18i \pi\alpha_s
 \bigg(
 \frac{2 m_{K^*} f_{\phi}^\perp }{m_b f_{\phi}} +
 \frac{2 m_{\phi} f_{K^*}^\perp }{m_b f_{K^*}} \bigg)   ( 2 X^-_A
-3)(X^-_A -1)\,.
 \end{eqnarray}

\section{New physics effects: $B\to h_1(1380) K^*$  vs $B\to \phi
K^*$}\label{sec:np}

In addition to annihilation contributions, the other possibility
for explaining the polarization puzzle in $B\to \phi K^*$ is to
introduce NP scalar- and/or tensor-like operators as discussed in
Ref.~\cite{Das:2004hq}. In the present paper, we will explore the
existing evidence in $B\to h_1(1380) K^*$ channel. The relevant NP
effective Hamiltonian ${\cal H}^{\rm NP}$, following the
definition in Ref.~\cite{Das:2004hq}, is given by
 \begin{eqnarray}
{\cal H}^{\rm NP}= \frac{G_F}{\sqrt{2}}\sum_{i=15-18, 23-26}
c_i(\mu) O_i (\mu)  + H.c.\,,
 \end{eqnarray}
where the scalar-type operators are
 \begin{eqnarray}
O_{15} &=& {\overline{s}}(1 + \gamma^5) b ~{\overline {s}}(1 +
\gamma^5) s \,,
  \ \ \ \
O_{16} =  {\overline{s}_{\alpha}} (1 + \gamma^5) b_{\beta}
  ~{\overline {s}}_\beta  (1 + \gamma^5) s_\alpha \,,
\nonumber \\
O_{17} &=&  {\overline{s}} (1 - \gamma^5) b
  ~{\overline {s}} (1- \gamma^5) s \,,
 \ \ \ \
O_{18} =  {\overline{s}_{\alpha}} (1 - \gamma^5) b_{\beta}
 s ~{\overline {s}}_\beta  (1 - \gamma^5) s_\alpha \,,
 \label{eq:scalarop}
 \end{eqnarray}
and the tensor-type operators are
 \begin{eqnarray} O_{23} &=&  {\overline{s}}\sigma^{\mu\nu} (1 + \gamma^5) b
  ~{\overline {s}}\sigma_{\mu\nu} (1 + \gamma^5) s \,,
 \ \ \ \
O_{24} =  {\overline{s}_{\alpha}}\sigma^{\mu\nu} (1 + \gamma^5)
b_{\beta}
  ~{\overline {s}}_\beta \sigma_{\mu\nu} (1 + \gamma^5) s_\alpha \,,
\nonumber \\
O_{25} &=&  {\overline{s}}\sigma^{\mu\nu} (1 - \gamma^5) b
  ~{\overline {s}}\sigma_{\mu\nu} (1 - \gamma^5) s \,,
 \ \ \ \
O_{26} = {\overline{s}_{\alpha}}\sigma^{\mu\nu} (1 - \gamma^5)
b_{\beta}
  ~{\overline {s}}_\beta \sigma_{\mu\nu} (1 - \gamma^5) s_\alpha \,,
   \label{eq:tensorop}
\end{eqnarray}
with $\alpha, \beta$ being the color indices.

 We now calculate the decay amplitudes for ${\overline
B^0} \rightarrow h_1(1380) {\overline K^{*0}}$ due to $O_{15-18}$
and $O_{23-26}$ operators defined in Eqs.~(\ref{eq:scalarop}) and
(\ref{eq:tensorop}). The CP-conjugate amplitudes for $B^0 \to
h_1(1380) K^{*0}$ can be obtained by $CP$-transformation. By the
Fierz transformation, $O_{15,16}$ and $O_{17,18}$  operators can
be expressed in terms of linear combination of $O_{23,24}$ and
$O_{25,26}$ operators, respectively, i.e.,
 \begin{eqnarray} \label{eq:fiez}
 O_{15} &=& \frac{1}{12}O_{23}- \frac{1}{6} O_{24} ,\ \ \
 O_{16} = \frac{1}{12}O_{24}- \frac{1}{6} O_{23} ,\nonumber
 \\
  O_{17} &=& \frac{1}{12}O_{25}- \frac{1}{6} O_{26} ,\ \ \
 O_{18} = \frac{1}{12}O_{26}- \frac{1}{6} O_{25} .
 \end{eqnarray}
In the computation, the matrix elements for tensor operators
$O_{23,25}$ can be recast into
\begin{eqnarray}
&& \langle h_1(1380)(q,\epsilon_{h_1}),\overline
K^*(p_{K^*},\epsilon_{K^*})|
 \bar s \sigma^{\mu\nu}  (1\pm \gamma_5) s\
 \bar s \sigma_{\mu\nu} (1\pm \gamma_5) b | \overline B(p) \rangle
  = \left(1+\frac{1}{2 N_c}\right) 8 f_{h_1}^\perp \nonumber\\
 && \ \  \times \Bigg\{
\pm i \epsilon_{\mu\nu\rho\sigma}
\epsilon_{h_1}^{*\mu}\epsilon_{K^*}^{*\nu} p_B^\rho
p_{K^*}^{\sigma} \, T_1(m_{h_1}^2) + \left[T_2(m_{h_1}^2)+
\frac{m_{h_1}^2}{m_B^2-m_{K^*}^2}T_3(m_{h_1}^2)\right]
(\epsilon_{h_1}^*\cdot p_B) \,(\epsilon_{K^*}^{*}\cdot p_B)
\nonumber\\
& &{ }\ \ \ \
 -\frac{1}{2} T_3(m_{h_1}^2) (\epsilon_{h_1}^*\cdot\epsilon_{K^*}^{*})
 \Bigg\}
\,, \label{eq:T}
\end{eqnarray}
under factorization, where the tensor decay constant
$f_{h_1}^\perp$ and the form factors  are defined
as~\cite{hfag,Ball:1998kk,kcymoments,Ball:1996tb}
 \begin{eqnarray}
  \langle h_1(1380)(q,\epsilon_{h_1})|{\overline{s} \sigma^{\mu\nu}\gamma_5 s}|0 \rangle
  = f_{h_1}^\perp (\epsilon_{h_1}^{\mu *} q^\nu - \epsilon_{h_1}^{\nu *} q^\mu ),
 \end{eqnarray}

 \begin{eqnarray}
 && \langle \overline K^*(p_{K^*},\epsilon_{K^*}) | \bar s
\sigma^{\mu\nu} \gamma_5 b | \overline B(p)\rangle = T_1(q^2)
[\epsilon_{K^*}^{\mu} (p_B+p_{K^*})^\nu -\epsilon_{K^*}^{\nu}
 (p_B+p_{K^*})^\mu]\nonumber\\
 &&\ \ \ + (T_1(q^2)-T_2(q^2))\frac{m_B^2-m_{K^*}^2}{q^2}
[ \epsilon_{K^*}^{\mu} q^\nu -\epsilon_{K^*}^{\nu} q^\mu)
]\nonumber\\
 &&\ \ \ -2\left[T_3(q^2) - (T_1(q^2)-T_2(q^2))\frac{m_B^2 -
 m_{K^*}^2}{q^2}\right]
  (p_{K^*}^\mu q^\nu - p_{K^*}^\nu q^\mu),
 \end{eqnarray}
 with
\begin{equation}
 T_1(0)  =  T_2(0). \label{eq:T1T2}
\end{equation}
The helicity amplitudes, in units of $G_F/\sqrt{2}$, for the
${\overline B^0}$ decay due to the NP operators are given by
 \begin{eqnarray} {\overline H}_{00}^{NP} &=&  4 f_{h_1}^\perp m_B^2 \left(
{\tilde a}_{23} + {\tilde a}_{25}\right)\left[h_2 T_2 (m_{h_1}^2)
- h_3 T_3(m_{h_1}^2)\right] \,,
\nonumber \\
{\overline H}_{\pm \pm}^{NP} &=& - 4 f_{h_1}^\perp m_B^2 \left[
\left(\frac{m_B^2-m_{K^*}^2}{m_B^2}\right)({\tilde a}_{23}+
{\tilde a}_{25}) T_2(m_{h_1}^2)  \mp 2p_c ({\tilde a}_{23} -
{\tilde a}_{25})T_1(m_{h_1}^2) \right] ,
 \end{eqnarray}
or in terms of the transversity basis,
 \begin{eqnarray}
  \label{npamp}
 {\overline A_{0}^{NP}} &=&   4 f_{h_1}^\perp m_B^2 \left(
{\tilde a}_{23} + {\tilde a}_{25}\right)\left[h_2 T_2 (m_{h_1}^2)
- h_3 T_3(m_{h_1}^2)\right] \,,
\nonumber\\
{\overline A_{\|}^{NP}} &=&  - 4 \sqrt{2} f_{h_1}^\perp
(m_B^2-m_{K^*}^2)
 ({\tilde a}_{23}+ {\tilde a}_{25}) T_2(m_{h_1}^2) ,
\nonumber \\
{\overline A_{\perp}^{NP}} &=& -  8 \sqrt{2} f_{h_1}^\perp m_B p_c
 ({\tilde a}_{23} - {\tilde a}_{25}) T_1(m_{h_1}^2),
 \end{eqnarray}
where
 \begin{eqnarray} \label{const}
h_2 &=& \frac{1}{2 m_{K^*} m_{h_1}}
\left(\frac{(m_B^2-m_{h_1}^2-m_{K^*}^2)(m_B^2-m_{K^*}^2)} {m_B^2}
- 4p_c^2 \right),
\nonumber \\
h_3 &=& \frac{1}{2 m_{K^*} m_{h_1}} \left(\frac{4p_c^2 m_{h_1}^2}{
m_B^2 - m_{K^*}^2} \right),
 \end{eqnarray}
and
 \begin{eqnarray}
{\tilde a}_{23} &=& a_{23} +\frac{a_{24}}{2}  - \frac{a_{16}}{8}\,, \nonumber\\
 {\tilde a}_{25} &=&  a_{25} +\frac{a_{26}}{2}  -
 \frac{a_{18}}{8}\,,
  \end{eqnarray}
are defined as ${\tilde a}_{23} = -\left|{\tilde a}_{23}\right|
e^{i\delta_{23}} e^{i\phi_{23}}, {\tilde a}_{25} = \left|{\tilde
a}_{25}\right| e^{i\delta_{25}} e^{i\phi_{25}}$ with
$\delta_{23,25}$ and $\phi_{23,25}$ being the corresponding the
strong phases and NP weak phases, respectively.\footnote{We have
restrained $|\delta_{23,25}|<\pi/2$.} Here we define
 \begin{eqnarray}
 a_{2i}&=&c_{2i}+ \frac{c_{2i-1}}{N_c}+{\rm nonfactorizable\ corrections}\, ,\nonumber\\
 a_{2i-1}&=&c_{2i-1}+ \frac{c_{2i}}{N_c}+{\rm nonfactorizable\ corrections},  \label{eq:wc}
 \end{eqnarray}
with $i \in$ integer number.

Since the $\phi K^*$ data showed that  $|\overline A_\perp|\, [=
-(\overline H_{++} - \overline H_{--})/\sqrt{2}] \simeq |\overline
A_\parallel| \, [= (\overline H_{++} + \overline
H_{--})/\sqrt{2}]$, there exist two possible solutions with
$\overline H_{++} \ll \overline H_{--}$ and  $\overline H_{++} \gg
\overline H_{--}$, respectively, for which the former, referring
to the NP scenario 2, corresponds to $|\tilde a_{23}|\simeq
1.5\times 10^{-4}, \tilde a_{25}\approx 0$, while the latter,
referring to the NP scenario 1, accords with $\tilde a_{23}\approx
0, |\tilde a_{25}|\simeq 2.0\times 10^{-4}$.

Comparing the NP results for $B\to h_1(1380) K^*$ with that for
$B\to \phi K^*$~\cite{Das:2004hq} (in units of $G_F/\sqrt{2}$):
 \begin{eqnarray} \label{npampphiKs}
 {\overline A_{0}^{NP}} &=& -4 i f_\phi^{\perp}
m_B^2 \left[ {\tilde a}_{23} - {\tilde a}_{25}\right] \left[h_2
T_2(m_\phi^2) -  h_3 T_3(m_\phi^2)\right] ,
\nonumber \\
{\overline A_{\|}^{NP}} &=&  4 i \sqrt{2} f_\phi^{\perp}
(m_B^2-m_{K^*}^2) ({\tilde a}_{23} - {\tilde a}_{25})
T_2(m_\phi^2) ,
\nonumber \\
{\overline A_{\perp}^{NP}} &=&  8 i \sqrt{2} f_\phi^{\perp} m_B
p_c ({\tilde a}_{23} + {\tilde a}_{25}) T_1 (m_\phi^2) ,
 \end{eqnarray}
we have  $|\overline A_{0}^{NP}(h_1(1380) K^*)/ \overline
A_{0}^{NP}(\phi K^*)| \simeq |\overline
A_{\parallel}^{NP}(h_1(1380) K^*)/ \overline
A_{\parallel}^{NP}(\phi K^*)| \simeq |\overline
A_{\perp}^{NP}(h_1(1380) K^*)/ \overline A_{\perp}^{NP}(\phi K^*)|
\simeq f_{h_1}^\perp/ f_{\phi}^\perp \approx 1$ for the case of
$\overline H_{++} \ll \overline H_{--}$ or $\overline H_{++} \gg
\overline H_{--}$. Neglecting the annihilation contributions, we
conclude that in the NP scenarios ${\rm BR_L}(h_1(1380) K^*)$ is
dominated by the SM amplitudes which may contain sizable QCD
corrections, while ${\rm BR_T}(h_1(1380) K^*)\approx (0.6 \sim
1.1) {\rm BR_T}(\phi K^*)$ due to the NP effects together with the
SM contributions in the $\phi K^*$ modes. The detailed analysis
will be given in the next section.

\section{Numerical results}\label{sec:results}

\subsection{Input parameters}
To proceed the numerical analysis, we adopt next-to-leading order
(NLO) Wilson coefficients in the naive dimensional regularization
(NDR) scheme given in \cite{BN}.  For Cabibbo-Kobayashi-Maskawa
(CKM) matrix elements, we adopt the Wolfenstein parametrization
with $A=0.801, \lambda=0.2265, {\overline \rho} = \rho (1 -
\lambda^2/2) = 0.189$ and ${\overline \eta} = \eta (1 -
\lambda^2/2) = 0.58$ \cite{PDG}. To take into account the possible
uncertainty of form factors on our results, in the numerical
analysis we combine two possible sets of form factors, coming from
the light-cone sum rule calculation. The form factors at zero
momentum transfer are cataloged in Table~\ref{tab:formfactors},
for which we allow 15\% uncertainties in values in the present
analysis, and their $q^2$-dependence can be found in
\cite{Ball:2004rg,Ball:1998kk}. The decay constants
\cite{Ball:2004rg,Ball:2003sc,Ball:1996tb} used in the numerical
analysis are collected in Table~\ref{tab:formfactors}. In analogy
with the QCD sum rule calculation for $f_{b_1}^\perp$, one can
obtain $f_{h_1}^\perp$~\cite{kcymoments}. We will simply take
$f_{b_1 (h_1)}=f_{b_1 (h_1)}^\perp(1~{\rm GeV})$ in the study
since only the products of $f_{b_1 (h_1)} a_1^{b_1
(h_1),\parallel}$ are relevant, where $a_1^{b_1 (h_1),\parallel}$
is the first Gegenbauer moment of $\Phi_\parallel^{b_1 (h_1)}$
defined in Eq.~(\ref{eq:phiparallel}). Using the QCD sum rule
technique, we have studied $a_1^{b_1 (h_1),\parallel}$ in
Ref.~\cite{kcymoments}, where the results are given by
 \begin{eqnarray}
 \begin{array}{lcl}
 a_1^{b_1, \parallel}(\mu=1~{\rm GeV})=-1.70\pm 0.45\,,
 & & \ \ a_1^{h_1,\parallel}(\mu=1~{\rm GeV})=-1.75 \pm 0.20\,, \\
 a_1^{b_1,\parallel}(\mu=2.2~{\rm GeV})=-1.41\pm 0.37\,,
 & &\ \  a_1^{h_1,\parallel}(\mu=2.2~{\rm GeV})=-1.45\pm 0.17\,. \\
 \end{array}
 \end{eqnarray}
The magnitudes of $a_1^{b_1 (h_1),\parallel}$ have a large impact
on the longitudinal fraction of the penguin-dominated  $B \to VA$
decay rates. We use the LCDAs of mesons given in
Eqs.~(\ref{eq:lcdas}), (\ref{eq:phiparallel}), and (\ref{eq:ww}).
It turns out that our predictions are insensitive to the 2nd
non-zero Gegenbauer moments of LCDAs. The integral of the $B$
meson wave function is parameterized as
 \begin{eqnarray}
 \int_0^1 \frac{d\rho}{\bar \rho}\Phi_1^B(\rho) \equiv
 \frac{m_B}{\lambda_B}\,,
 \end{eqnarray}
with $\lambda_B =(350\pm 150)$~MeV~\cite{BBNS1}. For simplicity,
the logarithmic divergences, $X_H, X_A$ are taken to be
independent of the helicities of the final states, with $\rho_H,
\rho_A\leq 1$ and $\phi_H, \phi_A \in [0,2\pi]$. As will be
discussed below, the values of $\rho_A$ and $\phi_A$ are further
constrained by the $\phi K^*$ data.
\begin{table}[htb]
 \centerline{\parbox{14cm}
{\caption{\label{tab:decayconstants} Values of decay constants
from QCD sum rule
calculations\cite{Ball:2004rg,Ball:2003sc,Ball:1996tb,kcymoments}}}}
\begin{ruledtabular}
\begin{tabular}{cccccc}
& $\rho$ &$K^*$ & $\phi$ & $b_1$ & $h_1(1380)$ \\
\hline
 $f_{V(A)}$ [MeV]
 &$205\pm 9$ & $217\pm5$ & $231\pm 4$ &$f_{b_1}^\perp(1~{\rm GeV})$
 & $f_{h_1}^\perp(1~{\rm GeV})$ \\
\hline\hline
 $f_{V(A)}^\perp(\mu=1~{\rm GeV})$ [MeV]
 &$160\pm 10$ & $170\pm10$ & $200\pm 10$ & $180\pm 10$ &  $200\pm20$ \\
\hline
 $f_{V(A)}^\perp(\mu=2.2~{\rm GeV})$ [MeV]
 &$147\pm 10$ & $156\pm 10$ & $183\pm 10$ &$165\pm 9$ &  $183\pm18$
\end{tabular}
\end{ruledtabular}
\end{table}
There are three independent renormalization scales for describing
the decay amplitudes: (i) $\mu_v$ for loop diagrams and and
penguin topologies, contributing to the hard-scattering kernels,
(ii) $\mu_H$ for the hard spectator scattering, and (iii) $\mu_A$
for the annihilation. We take $\mu_v\in [m_b/2, m_b]$ and
$\mu_{H,A}\in [1~{\rm GeV}, m_b/2]$~\footnote{It was argued in
Ref.~\cite{BBNS1} that $\mu_H, \mu_A \sim (\Lambda_h \mu)^{1/2}$
with $\mu \in [m_b/2, m_b]$. However, here we have taken larger
ranges of $\mu_H, \mu_A $ into account.}. The working scales,
$\lambda_B$, and values of form factors give a large impact on our
results. To reduce these theoretical uncertainties in predictions,
we constrain the parameters by means of $\overline B^0 \to \phi
\overline K^{*0}$ data. The relevant QCDF formulas for the $\phi
\overline K^{*0}$ mode can be found in
Refs.~\cite{Das:2004hq,Kagan}. Without the annihilation effects,
we illustrate the $\overline B^0 \to \phi \overline K^{*0}$
branching ratio corresponding to several typical choices of
parameters in Table~\ref{tab:phiks}, where since $X_H$ gives
corrections to $\overline H_{\mp\mp}$ and $\overline H_{00}$
suppressed by $1/m_b^{(2)}$ and $r_\chi^\phi$, respectively, the
results are insensitive to the magnitude of $\rho_H$. Four remarks
are in order. First,  the longitudinal fractions are $\gsim 85\%$
in Table~\ref{tab:phiks}. Second, we separately consider the
annihilation and new-physics effects. Third, the results in
Ref.~\cite{Kagan} indicate that if the annihilation corrections
construct to the negative polarization component (for $\overline
B$ decays), they become destructive to the longitudinal fraction
with the same order of magnitude. Since the data give ${\cal
B}(\overline B^0 \to \phi \overline K^{*0})=(0.95\pm 0.9)\times
10^{-6}$ and the longitudinal fraction $f_L=0.48\pm
0.04$~\cite{hfag}, it seems to be favored to have a larger value
($\gsim 0.8\times 10^{-6})$ of BR before adding the annihilation
effects, as some choices in Table~\ref{tab:phiks}; otherwise the
resulting branching ratio will be too small. (Thus form factors of
set 2 seem to be preferable.) If further considering the $\phi
K^*$ phase measurements with $1\sigma$ errors,
$\arg(A_{\parallel}/A_0)=2.36^{+0.18}_{-0.16}$ and
$\arg(A_{\perp}/A_0)=2.49\pm0.18$, we obtain $-45^\circ \lsim
\phi_A\lsim 10^\circ$. Forth, the new physics gives constructive
corrections to $A_0$, of order $1/m_b$, and to
$A_{\parallel,\perp}$, of order 1. Thus, to justify the
measurements, in the SM (without annihilation corrections), the
$\phi K^*$ BR should be $\lsim 4.5\times 10^{-6}$ before including
new-physics effects.

\begin{table}[tb]
\caption{Form factors at zero momentum transfer for $B\to V$
transitions evaluated in the light-cone sum rule analysis. Set~1
contains the newly results with some improvements in
Ref.~\cite{Ball:2004rg} and set~2 is the original analysis in
Ref.~\cite{Ball:1998kk}.} \label{tab:formfactors}
\begin{ruledtabular}
\begin{tabular}{cccccccc}
&$A_1$(0)&$A_2$(0)&$A_0$(0)&$V$(0)&$T_1$(0)&$T_2$(0)&$T_3$(0)\\
\hline
$B\to \rho$ (set~1) &0.242 & 0.221&0.303&0.323&0.267&0.267&0.176 \\
\hline
$B\to K^*$ (set~1)  &0.292 & 0.259&0.374&0.411&0.333&0.333&0.202 \\
\hline\hline
$B\to \rho$ (set~2) &0.261 & 0.223&0.372&0.338&0.285&0.285&0.202 \\
\hline
$B\to K^*$ (set~2)  &0.337 & 0.283&0.470&0.450&0.379&0.379&0.261 \\
\end{tabular}
\end{ruledtabular}
\end{table}

\begin{table}[tb]
\caption{The CP-averaged branching ratio (in units of $10^{-6}$)
for $\overline B \to \phi \overline K^{*0}$. The annihilation
corrections and possible new-physics effects are not included. The
longitudinal fractions are larger than 85\%. The results refer to
$\rho_H=0$, and the central values of decay constants and form
factors given in Tables~\ref{tab:decayconstants} and
\ref{tab:formfactors}. $\lambda_B$ is in units of MeV.}
\label{tab:phiks}
\begin{ruledtabular}
\begin{tabular}{c|ccc|ccc}
 & \multicolumn{3}{c|}{form factors: set 1}
 & \multicolumn{3}{c}{form factors: set 2} \\
 \hline
 & $\lambda_B=200$ & $\lambda_B=350$ & $\lambda_B=500$
 & $\lambda_B=200$ & $\lambda_B=350$ & $\lambda_B=500$ \\
  \hline
$\mu_v =m_b$   & 2.53 & 2.93 & 3.26 & 4.14 & 4.64 & 5.06  \\
$\mu_H =m_b/2$ &      &      &      &      &      &     \\
 \hline
$\mu_v =m_b/2$ & 6.19 & 6.82 & 7.34 & 9.81 & 10.6 & 11.2  \\
$\mu_H =m_b/2$ &      &      &      &      &      &     \\
 \hline
$\mu_v =m_b/2 $ & 2.84& 4.27 & 5.65 & 5.38& 7.33 & 9.12  \\
$\mu_H =1~{\rm GeV}$ &      &      &      &      &      &     \\
\end{tabular}
\end{ruledtabular}
\end{table}

\subsection{$B\to h_1(1380) K^*$}

To illustrate the nonfactorizable effects for
factorization-suppressed $B \to V A$ helicity amplitudes, where $A
(\equiv 1 ^1P_1)$ is formed by the emitted quarks from the weak
vertex, we give the numerical results for effective coefficients
$a_i^h$ in Table~\ref{tab:effectiveai}. The results are evaluated
at $\mu_v=\mu_H=m_b/2$, including all theoretical uncertainties.

In the SM, with parameters constrained by the $\overline B\to \phi
\overline K^*$ measurements and including the annihilation
effects, we have computed the branching ratios,  together with
relative phases among the amplitudes, of $h_1(1380) K^*$ modes,
which are summarized in Table~\ref{tab:h1ksann}. The QCD
corrections turn the local operators $\bar s\gamma_\mu \gamma_5 s$
into a series nonlocal operators and the resultant magnitudes of
the decay amplitudes depend on the first Gegenbauer moment
$a_1^{h_1,\parallel}$ of $\Phi_\parallel^{h_1}$. Unlike the case
of $\phi K^*$ modes, the two terms in the square bracket of
$A_3^{f,-}(h_1 \overline K^*)$ given in Eq.~(\ref{eq:Am}) are
mutually destructive such that the transverse (longitudinal and
perpendicular) BRs are less than $1\times 10^{-6}$ (see
Table~\ref{tab:h1ksann}). Nevertheless, the second term in the
square bracket of $A_3^{f,0}(h_1 \overline K^*)$ given in
Eq.~(\ref{eq:A0}) is much larger than the first one such that
annihilations contribute constructively to the longitudinal
amplitude which is thus remarkably enhanced. With annihilation
${\rm BR_L}(h_1 \overline K^{*0})$ could be $(9.0 \sim 16.1)\times
10^{-6}$.

Alternatively, if the large transverse component of the $\overline
B \to \phi \overline K^*$ branching ratio is due to the new
physics, then we expect that sizable transverse fractions can be
observed in $h_1(1380) \overline K^*$ modes. Without
annihilations, we have employed the experimental information on
polarization $\phi K^*$ decays~\cite{hfag} to determine the NP
parameter, $a_{23}$ or $a_{25}$, which characterizes our NP
scenario. Thus, in the NP scenario 1, we have $\tilde
a_{25}=|\tilde a_{25}| e^{i\delta_{25}} e^{i\phi_{25}}$ with
 \begin{eqnarray}\label{eq:a25}
 |\tilde a_{25}|=(2.0\pm 0.3)\times 10^{-4},\ \ \
 \delta_{25}=1.00\pm 0.30, \ \ \ \phi_{25}=-0.02\pm 0.06.
 \end{eqnarray}
 On the other hand, in the NP
scenario 2, we obtain $\tilde a_{23}=-|\tilde a_{23}|
e^{i\delta_{23}} e^{i\phi_{23}}$ with
 \begin{eqnarray}\label{eq:a23}
 |\tilde a_{23}|=(1.5\pm
0.3)\times 10^{-4}, \ \ \ \delta_{23}=-0.47\pm 0.20, \ \ \
\phi_{23}=-0.07\pm 0.06.
 \end{eqnarray}
Consequently, using the above $\tilde a_{23}$ and $\tilde a_{25}$
in the $h_1(1380) K^*$ modes, respectively, we show the results in
Table~\ref{tab:phiksnp}. Because the
transverse branching ratios 
are enhanced by NP operators, we therefore obtain sizable
transverse components: ${\rm BR_T}(h_1(1380) K^*)\approx (0.6 \sim
1.1) {\rm BR_T}(B\to \phi K^*)$. It should be stressed that,
unlike the case of $\phi \overline K^*$ modes, the two possible NP
solutions can be distinguished in the $h_1(1380) K^*$ modes since
there is no phase ambiguity existing between the two NP scenarios.

\begin{table}[htb]
\caption{Effective coefficients $a_i^{h(h')}$ for $B \to V A
(\overline B\to VA)$ helicity amplitudes, where $A [\equiv 1
^1P_1]$ is formed by the emitted quarks from the weak vertex,
obtained in the QCD factorization calculations. The results are
given at $\mu_v=\mu_H=m_b/2$.}\label{tab:effectiveai}
\begin{center}
\begin{tabular}{|c|c|c|c|c|c|c|}
\hline \hline
 $\Re(a^0_1)$ & $0.07\pm 0.02$  & $a^{+(-)}_1$ &$0.01\pm 0.01$
 & $a^{-(+)}_1$  & $0.02\pm 0.01$ \\
 $\Re(a^0_1)$ & $-0.04\pm 0.02$  & $a^{+(-)}_1$ &$-0.03\pm 0.02$
 & $a^{-(+)}_1$  & $-0.03\pm 0.02$ \\
\hline
 $\Re(a^0_2)$ & $-0.27 \pm 0.05$  & $a^{+(-)}_2$ &$-0.04\pm 0.02$
 & $a^{-(+)}_2$  & $-0.10\pm 0.03$  \\
 $\Im(a^0_2)$ & $0.16 \pm 0.04$  & $a^{+(-)}_2$ &$0.11\pm 0.03$
 & $a^{-(+)}_2$  & $0.11\pm 0.03$  \\
\hline
 $\Re(a^0_3)$ & $-0.012\pm 0.004$ & $a^{+(-)}_3$ &$0.002\pm 0.001$
 & $a^{-(+)}_3$  & $0.004\pm 0.002$  \\
 $\Im(a^0_3)$ & $-0.007\pm 0.003$ & $a^{+(-)}_3$ &$-0.005\pm 0.002$
 & $a^{-(+)}_3$  & $-0.005\pm 0.002$  \\
\hline
 $\Re (a^0_4)$ &$ -0.020\pm 0.005$ &$\Re(a^{+(-)}_4)$ &$-0.004\pm 0.002$
 & $\Re(a^{-(+)}_4)$& $-0.0001\pm 0.0001$  \\
 $\Im (a^0_4)$ &$- 0.012\pm 0.004$ &$\Im(a^{+(-)}_4)$ &$-0.001\pm 0.001$
 & $\Im(a^{-(+)}_4)$& $-0.0003\pm 0.0001$  \\
\hline
 $\Re(a^0_5)$ &$0.014\pm 0.003$   &$a^{+(-)}_5$  &$-0.007\pm 0.002$
 & $a^{-(+)}_5$  &$-0.004\pm 0.002$ \\
 $\Im(a^0_5)$ &$-0.009\pm 0.003$   &$a^{+(-)}_5$  &$0.006\pm 0.003$
 & $a^{-(+)}_5$  &$0.006\pm 0.003$ \\
 \hline
 $\Re(a^0_6)$ &$0.003\pm 0.001$ &$\Re(a^{+(-)}_6)$ & $0$
 & $\Re(a^{-(+)}_6)$ &$0$ \\
 $\Im(a^0_6)$ &$0.013\pm 0.005$ &$\Im(a^{+(-)}_6)$ & $0$
 & $\Im(a^{-(+)}_6)$ &$0$ \\
\hline
 $\Re(a^0_7)$ &$-0.0002\pm 0.0001$ &$\Re(a^{+(-)}_7)$ & $0.00008\pm 0.00003$
 & $\Re(a^{-(+)}_7)$ &$0.00004\pm 0.00002$ \\
 $\Im(a^0_7)$ &$0.00010\pm 0.00004$ &$\Im(a^{+(-)}_7)$ & $-0.00007\pm 0.00003$
 & $\Im(a^{-(+)}_7)$ &$-0.00007\pm 0.00005$ \\
 \hline
 $\Re(a^0_8)$ &$-0.00002\pm 0.00001$ &$\Re(a^{+(-)}_8)$ & $0$
 & $\Re(a^{-(+)}_8)$ &$0$ \\
 $\Im(a^0_8)$ &$0.00006\pm 0.00002$ &$\Im(a^{+(-)}_8)$ & $0$
 & $\Im(a^{-(+)}_8)$ &$0$ \\
\hline
 $\Re(a^0_9)$ &$-0.0006\pm 0.0002$  &$\Re(a^{+(-)}_9)$ &$-0.0001\pm 0.0001$
 & $\Re(a^{-(+)}_9)$ & $-0.0002\pm 0.0001$  \\
 $\Im(a^0_9)$ &$0.0003\pm 0.0002$  &$\Im(a^{+(-)}_9)$ &$0.0002\pm 0.0001$
 & $\Im(a^{-(+)}_9)$ & $0.0002\pm 0.0001$  \\
\hline
 $\Re(a^0_{10})$ &$0.0020\pm 0.0005$ &$\Re(a^{+(-)}_{10})$ &$0.0003\pm 0.0001$
 & $\Re(a^{-(+)}_{10})$& $0.0008\pm 0.0003$ \\
 $\Im(a^0_{10})$ &$-0.0014\pm 0.0004$ &$\Im( a^{+(-)}_{10})$ &$-0.0009\pm 0.0003$
 & $\Im(a^{-(+)}_{10})$& $-0.0009\pm 0.0004$ \\
\hline
\end{tabular}
\end{center}
\end{table}
\begin{table}[htb]
\caption{CP-averaged branching ratios (in units of $10^{-6}$) for
$\overline B \to h_1(1380) \overline K^*$ without/with
annihilation contributions denoted as BR$^{\rm wo}$/BR$^{\rm w}$.}
\label{tab:h1ksann}
\begin{ruledtabular}
\begin{tabular}{c|ccc|ccccc}
 & ${\rm BR_{tot}^{wo}}$ & ${\rm BR_{\parallel}^{wo}}$ & ${\rm BR_{\perp}^{wo}}$
 & ${\rm BR_{tot}^{w}}$ & ${\rm BR_{\parallel}^{w}}$ & ${\rm BR_{\perp}^{w}}$
 & $\arg(A_\parallel/A_0)$ & $\arg(A_\perp/A_0)$\\
\hline $ B^- \to h_1(1380)  K^{*-}$
 & $3.4^{+1.5}_{-1.2}$  & $\lesssim 0.2$ & $\lesssim 0.2$ & $13.1^{+4.1}_{-3.0}$
 & $\lesssim 1.0$ & $\lesssim 1.0$
 & $-0.72\pm0.13$ & $-0.71\pm 0.13$\\
 \hline
$ \overline B^0 \to h_1(1380)  K^{*0}$
 & $3.2^{+1.5}_{-1.2}$  & $\lesssim 0.2$ & $\lesssim 0.2$ & $12.0^{+4.1}_{-3.0}$
 & $\lesssim 1.0$ & $\lesssim 1.0$
 & $-0.67\pm0.13$ & $-0.67\pm0.13$
\end{tabular}
\end{ruledtabular}
\end{table}
{\squeezetable
\begin{table}[htb]
 {\caption{\label{tab:phiksnp} New-physics predictions
for $\overline B \to h_1(1380) \overline K^*$ modes, where BRs are
given in units of $10^{-6}$, and phases in radians. The input
parameters are used with constraints by the $B\to \phi K^*$ data.
$\tilde a_{25}$ and $\tilde a_{23}$ are given in
Eqs.~(\ref{eq:a25}) and (\ref{eq:a23}), respectively.}}
\begin{ruledtabular}
\begin{center}
 {
\begin{tabular}{c|c|ccccccc}
 New physics& Process & ${\rm BR_{tot}}$ & ${\rm BR_{\parallel}}$ & ${\rm BR_{\perp}}$
 & $\arg(\frac{A_\parallel}{A_0})$ & $\arg(\frac{\overline A_\parallel}{\overline A_0})$
 & $\arg(\frac{A_\perp}{A_0})$ & $\arg(\frac{\overline A_\perp}{\overline A_0})$\\
\hline
  Scenario 1: & $ B^- \to h_1(1380)  K^{*-}$
 & $15.3\pm 4.0$ & $3.4\pm 1.5$ & $2.1\pm 1.0$ & $-2.32\pm 0.25$ &
 $0.48\pm 0.15$ & $-2.21\pm 0.25$ & $0.49\pm 0.15$\\
 $\tilde a_{25}$& $ \overline B^0 \to h_1(1380)  K^{*0}$
 & $14.5\pm 4.0$ & $3.2\pm 1.5$ & $2.0\pm 1.0$ & $-2.23\pm 0.25$ &
 $0.49\pm 0.15$ & $-2.22\pm 0.25$ & $0.50\pm 0.15$\\
 \hline\hline
  Scenario 2: & $ B^- \to h_1(1380)  K^{*-}$
 & $9.1\pm 2.0$ & $2.1\pm 0.5$ & $2.0\pm 0.5$ & $-0.87\pm 0.20 $ &
 $-0.97\pm 0.20$  & $-0.87\pm 0.20$ & $-0.97\pm 0.20$\\
 $\tilde a_{23}$& $ \overline B^0 \to h_1(1380)  K^{*0}$
 & $8.5\pm 2.0$ & $2.0\pm 0.5$ & $1.8\pm 0.5$ & $-0.86\pm 0.20 $ &
 $-0.97\pm 0.20$  & $-0.87\pm 0.20$ & $-0.97\pm 0.20$
\end{tabular}}
\end{center}
\end{ruledtabular}
\end{table}
 }
{\squeezetable
 \begin{table}[htb] \caption{\label{tab:b1ksann}
CP-averaged branching ratios (in units of $10^{-6}$) for
$\overline B \to b_1^+(1235)  K^{*-}, \rho^+ \overline K^{*-}$
without/with annihilation contributions denoted as BR$^{\rm
wo}$/BR$^{\rm w}$.}
\begin{ruledtabular}
\begin{tabular}{c|ccc|ccccc}
 Process & ${\rm BR_{tot}^{wo}}$ & ${\rm BR_{\parallel}^{wo}}$ & ${\rm BR_{\perp}^{wo}}$
 & ${\rm BR_{tot}^{w}}$ & ${\rm BR_{\parallel}^{w}}$ & ${\rm BR_{\perp}^{w}}$
 & $\arg(\overline A_\parallel/\overline A_0)$ & $\arg(\overline A_\perp/\overline A_0)$\\
\hline $ \overline B^0 \to b_1^+(1235)  K^{*-}$
 & $1.7\pm 1.3$  & $\lesssim 0.01$ & $\lesssim 0.01$ & $7.0\pm 3.5$
 & $\lesssim 0.3$ & $\lesssim 0.3$
 & $-0.66\pm 0.15$ & $-0.66\pm 0.15$\\
 \hline
$ \overline B^0 \to \rho^+  K^{*-}$
 & $6.3\pm 2.0$  & $0.2\pm 0.1$ & $0.2\pm 0.1$ & $6.2\pm 2.0$ & $1.2\pm 0.7$ & $1.2\pm 0.7$
 & $2.18\pm 0.18$ & $2.17\pm 0.17$
\end{tabular}
\end{ruledtabular}
\end{table}
}

\subsection{$\overline B^0\to b_1^+(1235) K^{*-}\ vs.\ \overline B^0\to \rho^+ K^{*-}$}

The SM decay amplitudes for $\overline B^0\to b_1^+(1235) K^{*-},
\rho^+ K^{*-}$ read
\begin{eqnarray}
A_{\overline B^0 \to b_1^+ K^{*-}}^h &=&\frac{G_F}{\sqrt 2}
\Bigg\{V_{ub} V_{us}^* a_1^h X_h^{({\overline B} b_1, {\overline
K^{*}})} \nonumber\\
&& - V_{tb} V_{ts}^* \Bigg[ (a_4^h + r_\chi^{b_1} a_6^h +
r_\chi^{b_1} a_8^h + a_{10}^h ) X_h^{({\overline B} {\overline
K^{*}}, b_1)} + f_Bf_{K^*} f_{\rho} \bigg( b_3^h
 - \frac{1}{2}b_{\rm 3,EW}^h\bigg)\Bigg] \Bigg\},\nonumber\\
A_{\overline B^0 \to \rho^+ K^{*-}}^h &=&\frac{G_F}{\sqrt 2}
\Bigg\{V_{ub} V_{us}^* a_1^h X_h^{({\overline B} \rho, {\overline
K^{*}})} \nonumber\\
 && - V_{tb} V_{ts}^* \Bigg[
(a_4^h - r_\chi^{b_1} a_6^h - r_\chi^{b_1} a_8^h + a_{10}^h )
X_h^{({\overline B} {\overline K^{*}}, \rho)} + f_Bf_{K^*}
f_{\rho} \bigg( b_3^h
 - \frac{1}{2}b_{\rm 3,EW}^h\Bigg)\Bigg] \Bigg\},
\end{eqnarray}
respectively, where $a_i^h, b_i^h$ for $\overline B^0\to
b_1^+(1235) K^{*-}$ are given in Eqs.~(\ref{eq:ai}),
(\ref{eq:bi}), (\ref{eq:A0}), and (\ref{eq:Am}), while those for
$\overline B^0\to \rho^+ K^{*-}$ can be found in
Ref.~\cite{Das:2004hq}\footnote{$a_6^h$ and $a_8^h$ in
Ref.~\cite{Das:2004hq} should be corrected as in Eq.~(\ref{eq:ai})
of the present paper.} and Eqs.~(\ref{eq:bi}), (\ref{eq:XAA3}).
$b_1^+(1235) K^{*-}$ and $\rho^+ K^{*-}$ modes are
penguin-dominant processes. The former, only receiving the tiny
effect from the CKM suppressed tree amplitude for which the
longitudinal fraction is further suppressed by the $B\to b_1$
transition form factor~\footnote{\label{footnote:form} $A_1^{B
b_1}$ is expected to be much smaller than $A_1^{B \rho}$ since the
local axial vector current does not couple to $b_1$ in SU(2)
limit, and the local tensor current mainly couples to the
transverse states of $b_1$.}, is highly factorization-suppressed.
In the numerical study, we thus neglect the tree part of the
$\overline B^0\to b_1^+(1235) K^{*-}$ amplitude. Considering the
possible annihilation effects and using the input parameters also
constrained by the $B\to \phi K^*$ data, we have collected the
results for $\overline B^0\to b_1^+(1235) K^{*-}$ as well as for
$\overline B^0\to\rho^+ K^{*-}$ in Table~\ref{tab:b1ksann}. The
annihilation effects are negligible in ${\rm BR_T}({b_1^+
K^{*-}})$, but could give a significant enhancement of ${\rm
BR_L}(b_1^+ K^{*-})$. Nevertheless, for $\overline B^0\to\rho^+
K^{*-}$ the annihilation effects contribute constructively to the
transverse fractions with a large ratio of ${\rm BR_T}(\rho^+
K^{*-})/{\rm BR_{tot}}( \rho^+ K^{*-})=0.39^{+0.07}_{-0.16}$, but
destructively to the longitudinal component. It should be noted
that the contributions of NP tensor-type and scalar-type
four-quark operators to the transverse fractions of $\rho^+
K^{*-}$ and ${b_1^+ K^{*-}}$ modes are different from the cases of
$\phi K^*$ and $h_1 K^{*}$ modes. Since so far there is no data on
constraining these NP parameters, we do not further discuss this
possibility.

\subsection{$B^- \to b_1^- \rho^0$ and $\overline B^0 \to b_1^- \rho^+$}
The SM decay amplitudes with annihilation corrections for $B^- \to
b_1^- \rho^0$ and $\overline B^0 \to b_1^- \rho^+$ read
\begin{eqnarray}
A_{B^- \to b_1^- \rho^0}^h &\simeq&
 \frac{G_F}{\sqrt 2} \Bigg\{V_{ub}
V_{ud}^* ( a_1^h X_h^{({B^-} \rho^0, b_1^-)} +a_2^h X_h^{({B^-} b_1^-, \rho^0)}) \nonumber\\
&& - V_{tb} V_{td}^* \Bigg[ \frac{3}{2}(-a_7^h + a_9^h +
r_\chi^{b_1} a_8^h  + a_{10}^h ) X_h^{({B^-} \rho^0, b_1^-)} \Bigg] \Bigg\},\nonumber\\
A_{\overline B^0 \to b_1^- \rho^+}^h &\simeq&\frac{G_F}{\sqrt 2}
\Bigg\{V_{ub} V_{ud}^* \, a_1^h X_h^{({\overline B} \rho^+,
b_1^-)}
 - V_{tb} V_{td}^*\, \Bigg[(a_4^h + r_\chi^{b_1} a_6^h +r_\chi^{b_1} a_8^h + a_{10}^h )
X_h^{({\overline B} \rho^+, b_1^-) } \nonumber\\
&& + f_B f_{b_1} f_{\rho} \bigg( b_3^h - \frac{1}{2}b_{\rm 3,EW}^h
 \bigg)\Bigg]\Bigg\}.
\end{eqnarray}
Since $ X_h^{({B^-} b_1^-, \rho^0)}$ is negligible as explained in
footnote~\ref{footnote:form}, $\overline B^0 \to b_1^- \rho^0$ can
be roughly related to the tree dominated ${\cal B}(\overline B^0
\to \rho^- \rho^0)$ decay as
\begin{eqnarray}
 \frac{{\cal B}(\overline B^0 \to b_1^- \rho^0)}
 {{\cal B}(\overline B^0 \to \rho^- \rho^0)}
  \approx
  \left|\frac{f_{b_1} a_1^0 (b_1^- \rho^0)}
  {f_\rho a_1^0(\rho^- \rho^0)}\right|^2\approx 0.1^2.
\end{eqnarray}
Because the interference between the tree and penguin (including
annihilation) amplitudes is constructive in $\overline B^0 \to
b_1^- \rho^+$, we thus have a larger branching ratio for this
mode. Including annihilation contributions in the $b_1^- \rho^+$
mode, we obtain
\begin{eqnarray}
 {\cal B}(B^-  \to b_1^- \rho^0) &=& (0.3\pm 0.2)\times 10^{-6}, \nonumber\\
 {\cal B}(\overline B^0 \to b_1^- \rho^+) &=& (0.5\pm 0.3)\times
 10^{-6},
\end{eqnarray}
which are predominated by the longitudinal fraction.

\section{Conclusions}\label{sec:conclusion}

We have studied the factorization-suppressed $B$ decays involving
a $1 ^1P_1$ meson in the final state. For the penguin-dominated
$\overline B \to V\, V$ decays, the annihilation could give rise
to logarithmic divergent contributions $\sim {\cal
O}[(1/m_b^2)\ln^2 (m_b/\Lambda_h)]$ to the helicity amplitudes
$A_0$ and $A_{\parallel, \perp}$. Moreover, the annihilation
corrections interfere destructively and constructively in the
former and latter amplitudes, respectively. The branching ratios
for $\overline B^0 \to b_1^- \rho^+$ and $\overline B^0 \to b_1^-
\rho^+$ are $\lesssim 10^{-6}$. We show that if the large
transverse fractions of $\phi K^*$ mainly originate from the
annihilation topologies, then the large enhancement should be
observed only in the longitudinal component of $h_1 K^*$ and $b_1
K^*$ modes such that the resulting $f_L(h_1(1380) K^*)$ and
$f_L(b_1^+(1235) K^{*-})$ could be even larger than $f_L(\phi
K^*)$ and $f_L(\rho^+ K^{*-})$, respectively. Consequently, it is
interesting to note that ${\rm BR}(h_1(1380) K)$ and ${\rm BR}
(b_1^+(1235) K^{-})$ could be much larger than ${\rm BR}(h_1(1380)
K^*)$ and ${\rm BR} (b_1^+(1235) K^{*-})$,
respectively~\cite{gritsan}, since the annihilation effects are
further enhanced by the ``chirally-enhanced" factor. Roughly
speaking, we obtain

$$\frac{{\rm BR}(h_1(1380) K)}{{\rm BR}(h_1(1380) K^*)}\approx
 \frac{{\rm BR} (b_1^+(1235) K^{-})}{{\rm BR} (b_1^+(1235) K^{*-})}\approx 2.$$

On the other hand, if the large transverse fractions of $\phi K^*$
arise from the new physics, the same order of magnitudes of ${\rm
BR_T}(h_1 K^*)$ should be measured. Non-small strong phases of
$\tilde a_{23,25}$, as given in Eqs.~(\ref{eq:a25}) and
(\ref{eq:a23}), may hint at the (SM or NP) ${\it inelastic}$
annihilation topologies in decays. Although we do not
simultaneously take the annihilation and NP into account, the two
effects should be distinguishable in $h_1 K^{*0,-}$ and $b_1^+
K^{*-}$ modes.

According to the annihilation scenario, it is found that
$f_T(\rho^+ K^{*-})\simeq 0.23 \sim 0.46$. Analogously, one can
expect $2{\rm BR_T}(\rho^{0} K^{*-})\sim  {\rm BR_T}(\rho^{-}
\overline K^{*0})\sim {\rm BR_T}(\rho^+ K^{*-})$. Any obvious
deviation of the above relation from the experiments may imply the
new physics. It should be stressed that the $\rho^{0,+} K^{*-}$
and $\rho^- \overline K^{*0}$ modes are relevant for exploring NP
four-quark operators in $b \to s \bar u u$ and $b \to s \bar d d$
channels, respectively.

In analogy to the helicity discussion for $\overline B \to \phi
\overline K^*$ given in Ref.~\cite{Das:2004hq}, the helicity
structures of two-body baryonic $B$ decays were systematically
studied in Ref.~\cite{Suzuki:2005iq} based on the perturbative
argument. In the SM (even with considering the possible
annihilation effects), the dominant helicity amplitude is
$H_{-{1\over 2} -{1\over 2}}$ for the $\overline B$ decay. In
particular, it is interesting to note that $H_{+{1\over 2}
+{1\over 2}}$ could be remarkably enhanced in the NP scenario 1,
although $H_{-{1\over 2} -{1\over 2}}$ is dominant in the NP
scenario 2.

In summary, the measurements of $B\to h_1(1380) K^{*0,-}$ and
$b_1^+(1235) K^{*-}$ can offer a crucial test of our NP scenarios
and annihilation contributions.

\begin{acknowledgments}
 This work was supported in
part by the National Science Council of R.O.C. under Grant No:
NSC93-2112-M-033-004. I am grateful to Alex Kagan and Andrei
Gritsan for useful discussions.
\end{acknowledgments}

\end{document}